\documentclass{aa}

\usepackage{graphicx}
\usepackage{natbib}
\usepackage{scalerel}
\usepackage{amsmath}	% Advanced maths commands
\usepackage{multicol}        % Multi-column entries in tables
\usepackage{bm}		% Bold maths symbols, including upright Greek
\usepackage{pdflscape}	% Landscape pages
\usepackage[usenames,dvipsnames,table]{xcolor}
\usepackage{tabularx}
\usepackage{multirow}
\usepackage{multicol}
\usepackage{threeparttable}
\usepackage{txfonts}
\usepackage[pdfencoding=auto,psdextra]{hyperref}
\hypersetup{
    colorlinks=true,
    linkcolor=blue,
    filecolor=magenta,      
    urlcolor=blue,
    citecolor=blue
}
\makeatletter
\renewcommand*\aa@pageof{, page \thepage{} of \pageref*{LastPage}}
\makeatother

\usepackage[utf8]{inputenc}

\usepackage[switch, modulo]{lineno}
\def\Msun{\mbox{M$_\odot$}}

\def\mst{\mbox{$M_{\star}$}}

\def\lsim{\mathrel{\rlap{\lower3.5pt\hbox{\hskip0.5pt$\sim$}}
    \raise0.5pt\hbox{$<$}}}                % less than or approx. symbol
\def\gsim{~\rlap{$>$}{\lower 1.0ex\hbox{$\sim$}}}

\def\Fig{\mbox{Fig.~}}
\def\Figs{\mbox{Figs.~}}
\def\Tab{\mbox{Table~}}

\def\Sec{\mbox{Sect.~}}

\def\App{\mbox{Appendix~}}

\def\Rer{\mbox{$R_{\rm e,r}$}}

\def\Re{\mbox{$R_{\rm e}$}}

\def\sqd{\mbox{~sq.~deg.}}

\begin{document}
%
% Put the title and authors of your paper here
%

\title{VST-SMASH: VST Survey of Mass Assembly and Structural Hierarchy}
\subtitle{II. Exploring dwarf galaxies in the vicinity of NGC~5068 and of the two galaxies NGC~5084 and NGC~5087 at the edges of the Virgo Supercluster}
%I. Dwarfs and their companions NGC 5236/NGC 5253 and IC 5332/NGC 7713: tracing the mass assembly in the challenging faintest-end regime.
%% please do not edit the author list once you copy it from the
   
\newcommand{\orcid}[1]{} %% define as link to https://orcid.org/#1 if needed

\author{
C.~Tortora,$^{1}$\thanks{E-mail: crescenzo.tortora@inaf.it}
R.~Ragusa,$^{1}$
L.~Hunt,$^{2}$
A.~Unni,$^{1}$
M.~Baes,$^{3}$
Abdurro'uf,$^{4}$
F.~Annibali,$^{5}$
M.~Gatto,$^{1}$
N.~R.~Napolitano,$^{6}$
H.~Su,$^{6,7}$
A.~Venhola,$^{8}$
D.~Carollo$^{9}$
}

\institute{$^{1}$ INAF -- Osservatorio Astronomico di Capodimonte, Salita Moiariello 16, I-80131, Napoli, Italy\\
$^{2}$ INAF -- Osservatorio Astrofisico di Arcetri, Largo Enrico
Fermi 5, 50125, Firenze, Italy\\
$^{3}$ Sterrenkundig Observatorium, Universiteit Gent, Krijgslaan 281
S9, 9000 Gent, Belgium\\
$^{4}$ Department of Astronomy, Indiana University, 727 East Third Street, Bloomington, IN 47405, USA\\
$^{5}$ INAF -- Osservatorio di Astrofisica e Scienza dello Spazio di Bologna, Via Gobetti 93/3, I-40129 Bologna, Italy\\
$^{6}$ Department of Physics “E. Pancini”, University of Naples Federico II, Via Cintia, 21, 80126 Naples, Italy\\
$^{7}$ School of Mechanical, Electrical and Information Engineering, Shandong University,
180 Wenhua Xilu, Weihai, 264209, Shandong, China\\
$^{8}$ Space physics and astronomy research unit, University of Oulu,
Pentti Kaiteran katu 1, FI-90014 Oulu, Finland\\
$^{9}$ INAF -- Osservatorio Astronomico di Trieste, Via G. B. Tiepolo 11,
34143 Trieste, Italy
}

%\author{V. Busillo,$^{1,2,3}$\thanks{E-mail: valerio.busillo@inaf.it}
%C. Tortora,$^{2}$\thanks{E-mail: crescenzo.tortora@inaf.it}

%\institute{$^{1}$INAF -- Osservatorio Astronomico di Capodimonte, Salita Moiariello 16, I-80131, Napoli, Italy\\}

% 
% Put your abstract here:
%
\abstract{We present a study of dwarf galaxy candidates in the deepest optical imaging yet obtained of the field surrounding the nearby galaxies NGC~5068 (5.2 Mpc) and NGC~5084/NGC~5087 (25-30 Mpc), the latter two located in the peripheries of the Virgo Supercluster. This field, covering $\sim 2.6$~deg$^2$, was observed as part of the multi-band, wide-field and very deep data from VST Survey of Mass Assembly and Structural
Hierarchy (VST-SMASH), a distance-limited program ($D < 11$~Mpc) that reaches $g$- and $r$-band surface brightness depths of $\mu \sim 30$ mag arcsec$^{-2}$ within a box of $10 \times 10\, \rm arcsec^{2}$ at 1$\sigma$. Using a two-step visual inspection procedure, we identify 47 dwarf galaxy candidates
%
%, 24 secure and 18 uncertain, 
%
and perform the surface photometry of the sample and the fitting procedure with 1D Sérsic model on their profiles. Only four galaxies were previously reported in the literature, augmenting by one order of magnitude the number of dwarfs discovered in these regions. The colors (median $g-r = 0.57$ mag  and $r-i = 0.24$ mag) and structural properties of the dwarf candidates are consistent with the literature, as are their scaling relations with effective radius, Sersic index (n$< 2$), and absolute magnitude. We also investigate their central colour gradients, which exhibit significant scatter, and discuss them within the broader context of galaxy formation. We finally analyze the spatial distribution of these dwarf candidates relative to potential host galaxies. Considering projected separation, we identify reasonable associations with NGC~5084, NGC~5087, and NGC~5068 as likely hosts for a significant fraction of the sample. Several candidates are at physically credible distances from NGC~5068, despite what their offset size–luminosity relation alone might indicate. Future spectroscopic and deeper imaging follow-up is required to determine distances and velocities, enabling robust association with hosts, studies of satellite distributions and counts, and comparisons with cosmological expectations for planes of satellites and dark matter models.}
%
% Provide up to five key words:
%
%    from the list in
%     https://www.aanda.org/for-authors/latex-issues/information-files#pop}
%
% Add short versions of title and author list for page headings
%
   \titlerunning{VST-SMASH: paper II}
   \authorrunning{C. Tortora}
   
   \maketitle
%
%-------------------------------------------------------------------
%
%
%   Start the main text of your paper here
\section{Introduction}\label{sec:intro}

Dwarf galaxies represent the most abundant galaxy population in the Universe and are crucial testbeds for models of galaxy formation and evolution. Their low luminosities ($M_{\rm r} > -18$) and diffuse morphologies make them sensitive tracers of the interplay between baryonic and dark matter (DM) processes  (e.g., \citealt{Simon2007,Walker2013}). The abundance and spatial distribution of dwarf satellites have long been recognized as fundamental challenges for the standard $\Lambda$CDM paradigm. Even though current simulations have significantly improved, important aspects still need refinement, for example in the modelling of baryonic physics and in the implementation of alternatives to CDM that modify the DM flavour \citep{Bullock_Boylan-Kolchin17,Vogelsberger+20_simulations}. 

The absence of cored dark matter density profiles in DM-only $N$-body simulations, known as the cusp--core problem, was one of the earliest challenges to $\Lambda$CDM; however, modern hydrodynamical simulations that include baryonic physics and galaxy formation processes have largely alleviated this discrepancy \citep[e.g.,][]{Governato+12,Pontzen_Governato2012,Vogelsberger+20_simulations}. The so-called `missing satellites problem'---the discrepancy between the predicted abundance of DM subhalos and the comparatively small observed population of dwarf galaxies around the Milky Way (e.g., \citealt{Moore1999,Klypin1999,Willman2005,Bullock_Boylan-Kolchin17})---remains a matter of debate. However, some recent results suggest that it may no longer exist \citep[e.g.,][]{Lazar+25_no_missing_dwarfs_problem}, and the discovery of numerous ultra-faint dwarfs (UFDs) in wide-field surveys \citep{Koposov2015,homma2024PASJ...76..733H} instead seems to point to a `too many satellites problem'. In addition, the discovery of planar and corotating structures such as the Vast POlar Structure (VPOS) around the Milky Way \citep{Pawlowski2012,Pawlowski2015}, the Great Plane of Andromeda \citep{Ibata2013}, and similar structures around Centaurus A \citep{Muller+18_science} and NGC 4490/85 \citep{Karachentsev_Kroupa24} has presented further tension with cosmological expectations, as such kinematically coherent and flattened satellite distributions have been found to occur only rarely in $\Lambda$CDM simulations \citep{Pawlowski2014,Pawlowski+24,Cautun2015,Seo+24}. However, these inconsistencies are also a matter of debate, as other recent results instead suggest that such structures are not infrequent in the $\Lambda$CDM scenario \citep[e.g.,][]{Santos-Santos+20, Sales_Navarro23,Sawala+23_plane_of_satellites}. Finally, the diversity problem—which refers to the large observed variation in the inner rotation curves of galaxies with similar masses—was initially seen as being in strong tension with $\Lambda$CDM predictions, but has recently been largely alleviated \citep[e.g.,][]{Oman+15_diversity,Santos-Santos+20_diversity,Zentner+22_diversity,Cruz+25_diversity}.

To assess whether these issues are related only to the Local Group (LG) or represent a more universal phenomenon, it is necessary to extend dwarf galaxy surveys to nearby galaxy groups and isolated hosts in the Local Volume \citep[e.g.,][]{Muller2015,Muller+18_science}. Modern wide-field cameras have enabled systematic searches for faint, unresolved galaxies down to surface brightness (SB) levels of $\mu_{\rm r} \sim 29-30$ mag arcsec$^{-2}$, allowing the detection at distances of $\sim$4–5 Mpc of dwarfs with luminosities comparable to the classical LG satellites ($M_{\rm r} \sim -6$).

%at distances of $\sim$4–5 Mpc. 

In recent years, several deep imaging campaigns have significantly advanced our knowledge of the dwarf galaxy population in nearby clusters. 
In the Fornax cluster, both the Fornax Deep Survey \citep[FDS,][]{Venhola2018FDSIV,Venhola2019FDSVI,Venhola2022FDSXII} and the Next Generation Fornax Survey \citep[NGFS,][]{Munoz2015NGFS,Eigenthaler2018NGFS} have provided large, homogeneous catalogs of dwarf galaxies down to very low surface brightness (LSB) levels (with mean effective surface brightness $\overline{\mu}_{\rm e,r} \sim 26$ mag arcsec$^{-2}$), revealing their structural properties, nucleation fraction, and spatial distribution. 
Similar efforts have been carried out in the Virgo cluster, where the VST Early-type GAlaxy Survey  \citep[VEGAS,][]{Capaccioli2015VEGAS1,Spavone2017VEGAS2}, together with the Next Generation Virgo Cluster Survey \citep[NGVS,][]{Ferrarese2012NGVS,Prole2018VirgoVLSB}, have expanded the census of dwarfs and ultra-diffuse galaxies (UDGs), shedding light on environmental effects in dense environments such as rich clusters. VEGAS has conducted similar searches within the Hydra I cluster \citep{LaMarca+22_Hydra-I, LaMarca+22_Hydra-II}. Beyond these massive nearby clusters, the Assembly of early-Type Galaxies with their fine Structures survey \citep[MATLAS,][]{Duc2015MATLASOverview,Habas2020MATLASDwarfs,Poulain2021MATLASStruct} has mapped low-density fields and galaxy groups, uncovering a rich dwarf population and highlighting the role of less dense environment in shaping their morphology and nuclei. The LBT Imaging of Galaxy Haloes and Tidal Structures (LIGHTS) survey aims at reaching with LBT $\mu_{\rm V} \sim 31 \, \rm mag \, arcsec^{-2}$ in galaxies at distances $\sim 20$ Mpc \citep{Trujillo+21_LIGHTS}. Most recently, the \emph{Euclid} mission \citep{Mellier+25_Euclid}
has opened an entirely new window on dwarf galaxy studies in more distant groups, clusters and field environments. Its Early Release Observations (EROs) in the Perseus cluster revealed more than a thousand dwarf candidates \citep{Marleau2025PerseusERO}, while the first quick release (Q1) demonstrated \emph{Euclid}’s ability to compile large dwarf galaxy samples across a variety of environments \citep{Marleau2025Q1Census}. 
A first catalog of over 2600 candidates from \emph{Euclid} imaging further illustrates the transformative power of this mission for LSB science. These surveys provide a robust framework to study dwarf galaxy demographics and evolution across environments, from the nearby Universe to distant groups and clusters accessible with \emph{Euclid}.

Some work has also been dedicated to nearby galaxy groups such as the Centaurus group: \citet{Muller2015} reported the discovery of 16 new dwarf candidates within an area of $\sim$ 60 \sqd\ around the M\,83 subgroup, and subsequent work by \citet{Muller2017} extended this survey to $\sim500$ deg$^2$, uncovering an additional 41 candidates. These studies nearly doubled the known membership of the Centaurus complex. They highlighted the power of deep wide-field surveys in constraining the faint end of the galaxy luminosity function in environments beyond the LG. Complementary strategies have also been pursued. Ultra-deep targeted imaging with the Panoramic Imaging Survey of Centaurus and Sculptor (PISCeS) has revealed extremely faint satellites around Cen~A and NGC~0253, reaching down to $M_{\rm V} \sim -7$ \citep{Crnojevic2014,Crnojevic2016}. In the northern hemisphere, analogous work has led to the discovery of numerous dwarfs in the M\,81 and M\,101 groups \citep{Chiboucas2009,Merritt2014,Javanmardi2016}. Thanks to the use of machine learning, large-scale searches have been recently undertaken (e.g., \citealt{Paudel+23,Thuruthipilly+24_DES,Su+25_UDGs}). Together, these efforts provide critical benchmarks for testing cosmological models and understanding the environmental dependence of dwarf galaxy populations.

The VST -- Survey of Mass Assembly and Structural
Hierarchy (VST-SMASH) is designed to support this effort by providing deep, wide-field and multi-band optical imaging for a volume-limited sample of 27 nearby spiral galaxies ($D < 11$ Mpc), overlapping with the \emph{Euclid} Wide Survey in the Southern Hemisphere \citep{Tortora+24_VST-SMASH}. Building on previous surveys carried out with the VST \citep[e.g.,][]{Iodice2016ApJ...820...42I, spavone2017Galax...5...31S,ragusa2021A&A...651A..39R,ragusa2022FrASS...952810R,ragusa2023A&A...670L..20R}, VST-SMASH is conducted in the $g$, $r$, and $i$ band, to probe the mass assembly history of galaxies in the Local Volume, measuring the SB profiles of galaxies down to the faintest regimes (e.g. $\mu_{\rm g/r}\sim 30 \, \rm mag \, arcsec^{-2}$). This, in turn, allows to map LSB features and identify faint galaxies, such as UDGs and Ultra-Faint Galaxies (UFGs).

VST-SMASH has targeted the regions around the face-on spiral galaxy NGC~5068 at a distance of $\sim5.2$ Mpc \citep{Karachentsev+13}, and two members of the Virgo Supercluster, i.e., the massive edge-on lenticular galaxy NGC~5084 at $\sim$20.6~Mpc, and the elliptical galaxy NGC~5087 at $\sim$24.5~Mpc, to search for faint dwarf companions. 
These targets offer an ideal field for extending the previous investigations, in a region which is overlapping with the peripheries of the Virgo supercluster, and is not covered by recent wide-field searching campaigns \citep[e.g.,][]{Paudel+23,Thuruthipilly+24_DES}. We aim to constrain the satellite systems of NGC~5068 and NGC~5087/5084 and assess whether their dwarf galaxy populations follow the typical trends observed in the literature. This study contributes addressing the apparent tensions between observations and $\Lambda$CDM predictions.

The paper is organized as follows. In \Sec\ref{sec:data}, we introduce VST-SMASH and describe the observations and data reduction. The search for dwarf galaxy candidates and the procedures used to obtain the SB profiles in each band,  measure their colors, and structural parameters are presented in \Sec\ref{sec:search_characterize}. In \Sec\ref{sec:results}, we discuss integrated colors, structural parameters, colour gradients, and scaling relations, together with the spatial distribution of the candidates around to potential host galaxies. Finally, our conclusions are summarized in \Sec\ref{sec:conclusions}. Through the paper, all magnitudes are given in the AB system \citep{Oke_Gunn83}  and are corrected for Galactic extinction, adopting values of 0.339, 0.234, and 0.174 mag in the $g$, $r$, and $i$ bands,\footnote{\url{https://ned.ipac.caltech.edu/cgi-bin/nph-objsearch?objname=NGC+5068&img_stamp=YES&list_limit=9&extend=no##GalacticExtinction_0}.} respectively \citep{schlafly11}.

\section{Observations and data reduction}\label{sec:data}

VST-SMASH aims to achieve a comprehensive characterisation of stellar populations, tidal features, dwarf galaxies, globular clusters, and other LSB structures around galaxies within a volume-limited distance of approximately 10 Mpc. The survey has been introduced in \cite{Tortora+24_VST-SMASH} and \cite{Ragusa+26_VST-SMASH}; we refer the reader to these works for a detailed overview, while here we focus on the specific observations of the VST-SMASH field analysed in this study. Leveraging the capabilities of the VST and OmegaCAM, as demonstrated in \cite{Ragusa+26_VST-SMASH} the survey reaches SB limits of $\sim$ 30, $\sim$ 30, and $\sim$ 28 mag/arcsec$^2$ in the $g$, $r$, and $i$ band, respectively.

The field analysed here comprises two contiguous VST pointings, covering NGC~5068—one of the 27 primary survey targets—and an adjacent field used primarily to estimate and subtract the background level around NGC~5068. These regions, however, partially overlap with the peripheries of the Virgo Supercluster, including several member galaxies. In particular, the field contains NGC~5084, a large edge-on lenticular galaxy at  $D=20.6$~Mpc, NGC~5087, an elliptical galaxy at $D=24.5$~Mpc, and other smaller members of the Supercluster. According to the VST-SMASH observing strategy, each field is observed through five pointings in the $g$ and $r$ bands and four pointings in the $i$ band, centered on different positions. For each pointing, a standard diagonal dithering strategy is adopted, consisting of six short exposures of 300\,seconds each. Under the nominal survey strategy, this corresponds to total integration times of 2.5\,hours in the $g$ and $r$ bands and 2\,hours in the $i$ band, while efficiently bridging the gaps between the 32 CCDs of OmegaCAM. For the field analysed in this work, additional observations were obtained, resulting in total integration times of approximately 2.7, 3.4, and 2.2\,hours in the $g$, $r$, and $i$ bands, respectively.

Observations were carried out during the VST-SMASH runs 110.25AA.001 (Mar–Apr 2023), 112.266Z.001 (Mar 2024), and 113.26YY.001 (Mar 2024), resulting in a mosaic covering 2.69, 2.70 and 2.65~\sqd\ in $g$, $r$ and $i$ bands respectively. 

In this paper, we focus on the identification of dwarf galaxies candidates and on their photometric and structural characterisation. We will analyze the surface photometry of NGC~5068 and the main Virgo Supercluster members in future works.

Data reduction and calibration were performed using the AstroWISE pipeline, specifically developed for OmegaCAM observations \citep{McFarland2013ExA....35...79M, McFarland2013ExA....35...45M}. The pipeline includes bias and flat-field corrections, sky background subtraction, and the creation of the co-added mosaic \citep[for further details, see][]{venhola17}, with a final spatial scale of 0.2 arcsec per pixel.

The point-spread function (PSF) full width at half maximum (FWHM) is $\sim 1.1$, $0.8$, and $1.0$~arcsec\footnote{At the distances of NGC~5068, NGC~5084, and NGC~5087, this scale corresponds to 
$\sim 0.025$, $0.10$, and $0.12 \,\mathrm{kpc\,arcsec^{-1}}$, respectively.} in the $g$, $r$, and $i$ bands, respectively\footnote{The Python routine {\tt DAOStarFinder} is used to identify stars within the fields around the selected dwarf candidates, which are then fitted with a Moffat profile. We report the median FWHM values.}. The 1$\sigma$ image depths within a region of 100 arcsec$^2$, calculated following \cite{Hunt+25_Showcase} and \cite{Ragusa+26_VST-SMASH}, are 30.7, 30.2 and 29.4 mag\,arcsec$^{-2}$ in $g$, $r$, and $i$, respectively, confirming the effectiveness of our observing strategy and meeting the goals of the survey\footnote{They correspond to 29.5, 29 and 28.2 mag\,arcsec$^{-2}$ image depths at 3$\sigma$.}.

\begin{figure*}
\centering
\includegraphics[width=0.9\linewidth]{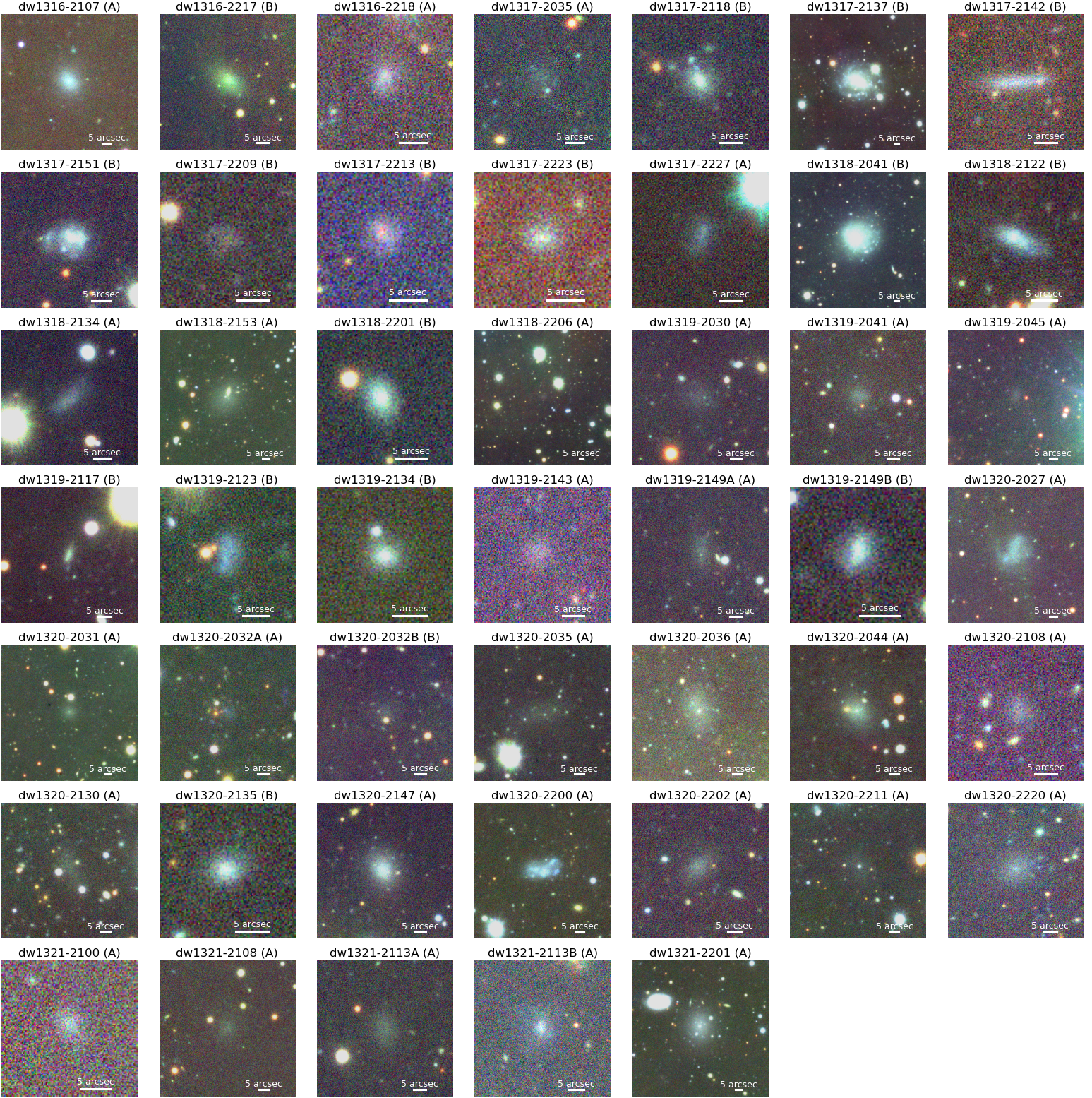}
\caption{RGB images of the selected dwarf candidates after the tagging and vetting stages of our visual classification. North is up and East is to the left. The horizontal white bar corresponds to 5 arcsec. Each cutout has a size equal to 10 times the effective radius in the $r$ band  of the galaxy.
The assigned vote is given in parentheses.}
\label{fig:rgb_dwarfs}
\end{figure*}

\section{Search and characterisation of dwarf candidates}\label{sec:search_characterize}

In this section, we describe the procedure for selecting dwarf galaxy candidates, which was carried out in several steps. We began with a tagging procedure to identify potential candidates, followed by a vetting process to assess the likelihood of their dwarf nature, and finally, after careful masking, measurements of integrated photometry and structural parameters.

\subsection{Tagging and vetting stages}

Our classification was carried out in two stages: first, a `tagging' stage, aimed at a preliminary labeling of the targets; followed by a `vetting' stage, in which each object was analyzed in greater detail and assigned a quality grade.

For the tagging stage, we did not run any automatic detection algorithm, as the faint objects of interest could easily be missed. Instead, following previous approaches by other studies (e.g., \citealt{Marleau2025Q1Census}) and given the relatively small field of view (FoV), we conducted a visual inspection. Unlike other searches, we did not rely on any preselection criteria, but instead examined the full mosaic to identify potential dwarf galaxy candidates, inspecting the $g$-band image. The final sample was constructed joining the visual inspection results from three co-authors (C.T., R.R. and M.B.). We searched primarily for diffuse, “fluffy” galaxies with clear low surface brightness, excluding obvious background spiral galaxies and systems with visual steep light profiles. However, we retained a few objects that are clearly nearby and exhibit some structure (e.g., spiral arms), such as low-mass spirals resembling IC 5332 \citep{Ragusa+26_VST-SMASH}; this is the case for the galaxy ESO~576-23, which is included in our final dwarf sample (see below). Although we aimed to avoid very small objects that are more likely to be background galaxies, we adopted a conservative approach, including all sources based on relatively relaxed visual size cuts. This procedure resulted in a preliminary sample of 71 dwarf candidates.

Each target was enclosed within a circular aperture with a radius at least as large as the visible extent of the galaxy. Cutouts for the visual inspection in the vetting stage were then generated, with sizes set to five times the chosen circle diameter to facilitate the inspection. RGB images were constructed from the $g$-, $r$-, and $i$-band cutouts using the {\it Python} routine {\tt make\_lupton\_rgb} \citep{Lupton+04}. At this stage, we reassessed the nature of the candidates incorporating the ($g - r$ and $r - i$) colour information, from both the images and aperture-based measurements, in the criteria adopted during the tagging phase. In this way we excluded extremely blue or red objects. The three inspectors independently examined the cutouts and assigned a grade to each candidate: A (high-probability dwarf), B (likely dwarf), or X (not a dwarf). We assign the final grade as the average between the three independent assessments.\footnote{We assign the B vote if the three grades were different, and the majority vote in case at least two inspectors voted with the same score.} Following this vetting stage, a sample of 47 dwarf candidates was selected: 30 grade A and 17 grade B. The RGB images of the selected candidates are shown in \Fig\ref{fig:rgb_dwarfs}. The corresponding $r$-band images are shown in \Fig\ref{fig:r-band_images}.

\subsection{Automatic and manual masking}

To perform both integrated and surface photometry, it is essential to mask foreground stars and other contaminants, given the faintness of our targets. This procedure was carried out in two stages. The first stage is fully automatic: stars (and background galaxies) are detected in the $r$ band using the {\it Python} routine {\tt DAOStarFinder}, with a detection threshold set to five times the background rms, estimated in the outer regions of the processed images (sufficiently far from the targets). Circular masks were then applied to the detected sources in each band, with radii scaled according to the source flux.

In the second stage, we performed a complementary manual masking to refine the process, especially in the regions surrounding each dwarf candidate and for the faintest targets. In particular, we enlarged the masks around saturated stars, masked elongated galaxies that were missed by the automatic routine, and added masks for extremely faint and compact background sources, as well as any remaining image defects. The masked $r$-band images are shown in \Fig\ref{fig:masked_r-band_images}.

\subsection{Integrated and surface photometry}

We first measured aperture photometry using a fixed circular aperture and then derived structural parameters, including aperture magnitudes, by fitting a S\'ersic law to the 1D SB profile calculated within circular annuli. All photometry was performed in each band on the masked images.\footnote{We developed our custom code in \texttt{Python}, employing the following key packages: \texttt{Astropy} for handling FITS data and visualizing images, \texttt{Photutils} for star detection and aperture photometry, and the function \texttt{curve\_fit} of \texttt{SciPy} for Sérsic profile fitting.}

Therefore, as a first step, we obtained a preliminary estimate of the magnitudes by computing the integrated flux within a circular aperture with a radius of 5 arcsec, corresponding to a diameter ten times the average FWHM across all bands. The associated errors were computed combining the weighted maps produced by AstroWISE and poisson noise. The following formula is used for the error $\sigma = \sqrt{\sigma_{\mathrm{sky},i}^{2} + \frac{f_i}{\mathrm{GAIN}}}$, where $\sigma_{\mathrm{sky},i} = \frac{1}{\sqrt{W}}$, with $W$ being the pixel value in the weight image, $f_{i}$ the flux in the corresponding pixel of the science image, and $\mathrm{GAIN}$ the ratio of the calibrated flux units to observed electrons, as calculated by SWarp during the production of the mosaic images.\footnote{Weight maps encode both the expected noise per pixel and the location of bad or contaminated pixels. Hot and cold pixels identified from the bias and flat-field images are set to zero in the weight maps. The weight maps are thus defined as $W=M_{\rm bad}/\sigma^2$, where $\sigma$ is the combined standard deviation of sky background and the read-out noise and $M_{\rm bad}$ is the bad-pixel mask (0 for bad pixels, 1 otherwise).} This fixed aperture was chosen  to provide an initial, model-independent estimate of the flux, sufficient for deriving preliminary magnitudes and performing quality checks. Final magnitudes are obtained using more accurate, model-dependent methods based on S\'ersic fitting, and we focus on magnitudes calculated within the effective radius, as described in the following paragraph. 

We performed SB photometry, adopting a more refined approach. To ensure a consistent reference across all filters, we performed a weighted co-addition of the $g$, $r$, and $i$ bands to create a deep detection image. This approach maximizes the S/N to find a common center, which is essential to derive reliable color profiles and avoiding artificial gradients caused by centering offsets between different bands. The center on the detection image is searched within a $10 \times 10$ pixel box (i.e., $2 \times 2 \, \rm arcsec^{2}$) around the initial center (visually identified), and assigning the final center to the location with the highest median flux in a $3 \times 3$ pixel sub-box. 

Radial SB profiles were extracted in elliptical annuli with fixed ellipticity and position angle (PA). The {\tt Source Extraction and Photometry (SEP, \citealt{Barbary+18_sep})} library was used on the detection image to determine the shape parameters, and it successfully provided reliable estimates for most galaxies in the sample. For the remaining faint or morphologically complex systems, the parameters were determined visually; a circular aperture was adopted when no clear elongation was detected from the visual inspection. The median ellipticity of the sample is 0.23, with a $1\sigma$ interval of 0.02--0.42. The radial profiles were sampled using an initial central bin width of 2 pixels ($\sim 0.4$ arcsec), which subsequently increases by a geometric growth factor of 1.1. This configuration preserves high spatial resolution within the central regions (sampling at roughly half the FWHM) while significantly enhancing the $S/N$ in the outskirts by averaging over progressively larger areas as the surface brightness declines. The flux in each annulus was computed as a weighted mean of the pixel values, with the weight maps providing the weights and the corresponding uncertainty calculated accordingly, also including the contribution from the Poisson noise as described above. A background level, determined as the median flux in annuli sufficiently far from the galaxy 
was subtracted from the annular fluxes. This procedure yields both 1D SB and $S/N$ profiles, which are truncated at the radius where $S/N < 3$ in each band. Integrated magnitudes were then derived by integrating the SB profiles out to the chosen radius. 

Defining the profile depth as the SB value corresponding to the 
outermost point of each profile, we find median values of SB (S/N) of 28.1 (3.8), 28.1 (3.8), and 27.1 (3.8) mag arcsec$^{-2}$, corresponding to median outermost 
radii of 9.1, 10.5, and 10.5~arcsec in the $g$, $r$, and $i$ bands, respectively. 
Even after accounting for differences in the S/N definition and aperture area, the 
depths derived from the profiles remain systematically shallower  than those 
calculated using empty apertures (of $0.4-0.8$ mag arcsec$^{-2}$). This persistent discrepancy is driven primarily by the differing noise properties: 
empty-aperture measurements rely purely on the Poisson noise of the blank sky 
background, whereas radial profile bins encompass the extended light of the target 
galaxy. Even in the faint outskirts, the source photons introduce their own 
intrinsic Poisson noise, which adds to the sky background and inflates the total 
variance. Furthermore, azimuthal profile extraction is more sensitive to 
localized background variations, residual flat-fielding asymmetries, and undetected 
faint background sources trapped within the extraction annuli. These combined 
factors naturally increase the measured noise floor in the profiles, resulting in a shallower effective depth compared to measurements taken in pristine, empty sky 
regions. However, when the profiles are recomputed using only the sky background errors, and after correcting the reference depths for the different S/N thresholds and aperture sizes, we find that the two depth estimates agree within $\sim 0.2\, \rm mag \, arcsec^{-2}$.

The 1D SB profiles were then fitted with a single-component Sérsic model to derive the best-fit structural parameters\footnote{For the sake of uniformity and because some galaxies are faint, we opted to use our home-made Python code to fit a Sérsic profile to the 1D surface brightness profile rather than fitting a 2D Sérsic profile directly to the image.}:  the effective SB $\mu_{\rm e}$, the effective radius \Re, and $n$, from which $\mu_{0}$ and the total Sérsic magnitude were also inferred. Although most dwarf galaxy candidates' profiles are expected to be shallow (low Sérsic index, e.g., $n \lsim 2$) with effective radii larger than the FWHM, before fitting, the Sérsic profile was convolved with a Gaussian function matching the FWHM of each band. The 1D SB profiles, together with the PSF-convolved and unconvolved Sérsic fits,\footnote{OmegaCAM exhibits extended, exponential-like PSF wings. Following \citet{Capaccioli+15} and Fig. G.1 of \citet{Ragusa+26_VST-SMASH}, we find that the peak of the wing is $\sim 2 \times 10^{5}$ times fainter than the PSF core peak, demonstrating that its contribution is negligible and that modelling only the core is a reasonable approximation.} are shown in \Fig\ref{fig:SB_profiles}.

\begin{figure}
\centering
\includegraphics[width=0.99\linewidth]{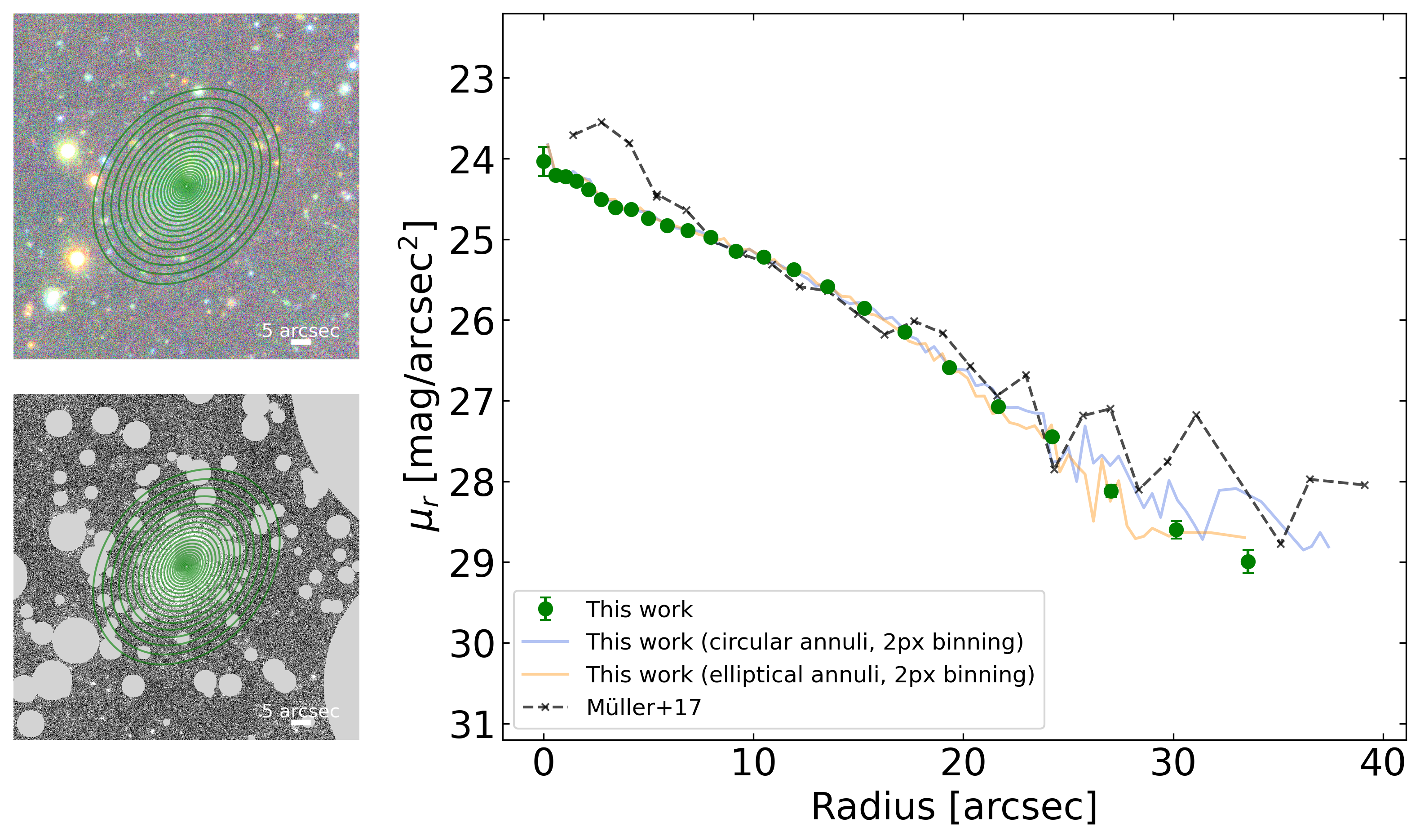}
\caption{Surface brightness profile for dw1318$-$2153. Left panels: RGB color-composite image (top) and masked $r$-band image (bottom). The green ellipses correspond to the geometric mean of the bin intervals used to derive the SB profile shown in the right panel. Masked regions are highlighted in light gray. Right panel: Extinction-corrected $r$-band SB profile derived in this work (green circles with error bars) compared with the results from \citet[][cross markers with dashed line]{Muller2017}. Profiles obtained with a fixed 2-pixel binning in both elliptical (orange) and circular (blue) apertures are also shown.}
\label{fig:dw1318-2153_profile}
\end{figure}

\begin{figure*}
\centering
\includegraphics[width=0.95\linewidth]{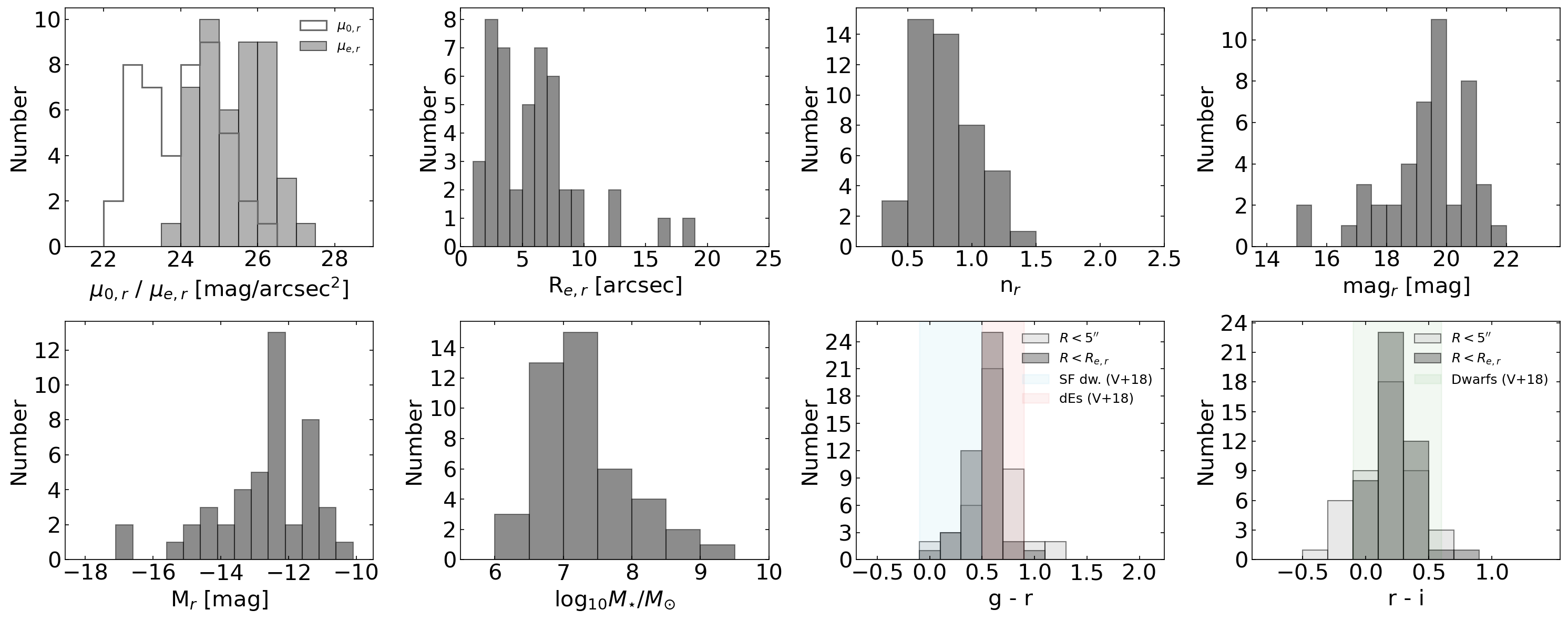}
\caption{Distributions of the physical properties of the dwarf galaxy candidates. From top-left, the panels show the following quantities: central surface brightness ($\mu_{\rm 0,r}$) and effective surface brightness ($\mu_{\rm e,r}$), circularized effective radius (\Rer), Sérsic index ($n_{\rm r}$), Sérsic total magnitude ($mag_{\rm r}$), Sérsic absolute magnitude ($M_{\rm r}$), total stellar mass ($\log_{10} \mst / \Msun$, see how it is calculated in \Sec\ref{subsec:phypar}), and the $g-r$ and $r-i$ colours. All Sérsic parameters shown are measured in the $r$ band. Magnitudes, SBs, and colours are extinction-corrected. In the last two panels, shaded regions indicate the approximate colour ranges of dwarf candidates from \cite{Venhola2018FDSIV,Venhola2019FDSVI}: star-forming dwarfs (cyan) and dwarf ellipticals (red) in the panel for $g-r$ color, and the general dwarf population (green) in the panel for $r-i$ color. We are only showing galaxies with relative errors on the $r$-band $R_{\rm e}$ and $n$ smaller than 50\%; the only excluded galaxy is dw1318-2206.}
\label{fig:histograms}
\end{figure*}

\section{Results}\label{sec:results}

Inspection of the $g$-band and RGB images (e.g. \Fig\ref{fig:rgb_dwarfs}) shows that most galaxies exhibit a smooth morphology, typical of dwarf ellipticals or dwarf spheroidals. A minority (about 10) display internal features such as arms, blobs, or central overdensities.\footnote{We caution the reader that the coarse resolution from the Earth can smooth existing substructures.} 

We now explore their structural parameters, colours, and their spatial distribution with respect to that of possible host galaxies. General information on the candidates, together with some photometric measurements, is reported in \Tab\ref{tab:tab1}, while structural parameters are listed in \Tab\ref{tab:tab2}. The names of the galaxies are created by prefixing ‘dw’ to their RA (hours and minutes) and Dec (degrees and minutes) coordinates.

\subsection{A literature cross-check}

To our knowledge, all but four of these candidates have not been previously reported in the literature. We searched for cross-matches in SIMBAD and NED database, and used the Heraklion Extragalactic CATaloguE (HECATE, \citealt{Kovlakas+2021_HECATE}) to complement our data with additional information. HECATE is a catalog of nearby galaxies within $\sim200$ Mpc, based on HyperLEDA \citep{Makarov+14_hyperleda},\footnote{http://leda.univ-lyon1.fr/} and supplemented with data from other extragalactic, photometric catalogs. It provides distances derived from a combination of redshifts and redshift-independent indicators.

The four previously known galaxies are described in the following.

\begin{itemize}
    \item ESO~576-23 (dw1317-2137) is identified as an Active Galactic Nucleus (AGN) candidate in SIMBAD, with a redshift of $\sim$ 0.009819 \citep{Meyer+04_HIPASS-I}, and a redshift-based distance of $\sim$ $39.9 \pm 6.9$ Mpc in the HECATE catalog. From our images, it appears to be a small, irregular/spiral galaxy;
    \item ESO~576-25 (dw1318-2041) is classified as an AGN candidate in SIMBAD, with a morphological type of -5, and a redshift-based distance of $\sim$ $22.4 \pm 4.6$ Mpc in the HECATE catalog. This is an irregular galaxy, resembling in its shape Sextans A;
    \item [CCF97]G4 or GALEXASC~J132012.36-220008.9 (dw1320-2200) is also classified as an AGN candidate in SIMBAD, with a redshift of $\sim$ $0.003643 \pm 0.000410$ \citep{Carignan+97_NGC5084} and a distance of $\sim$ $24.0 \pm 4.8$ Mpc in the HECATE catalog. This galaxy has an irregular morphology, featuring several distinct bright clumps, and is the faintest galaxy of the four;
    \item dw1318-2153 was reported as a new dwarf candidate (dw1318-21) in \cite{Muller2017}. This galaxy exhibits a smooth morphology, typical of dwarf ellipticals or dwarf spheroidals.
\end{itemize}

Since dw1318-2153 is the second faintest of the four systems, it provides a useful benchmark to compare our results with those of a similar search. \cite{Muller2017} observed a wide area of the sky using the Dark Energy Camera at the 4-m Blanco telescope, with a seeing of $\sim 1$ arcsec. Based on the galaxy's location within their footprint and the numbers reported in \cite{Muller2017}, the field around this galaxy would have been observed with exposure times in the ranges 240-420 s in the $g$ band and 200-340 s in the $r$ band. Scaling for telescope diameter and exposure time, their observations would need roughly one hour of exposure to reach the depth of our VST observations; our images are therefore about one magnitude deeper. This likely explains why some of our galaxies are missing in their catalog.

The extinction-corrected $r$-band SB profile of dw1318$-$2153 is compared with the results from \citet{Muller2017} in the right panel of \Fig\ref{fig:dw1318-2153_profile}. On the left, we also show the RGB color-composite (top) and masked $r$-band (bottom) images, overlaid with the ellipses corresponding to the geometric radii used in the SB profile. While we find excellent agreement in the $7\text{--}15$~arcsec range, some discrepancies emerge at smaller and larger radii. In particular, an unphysical rise at $R \lesssim 5$~arcsec is visible in the \citet{Muller2017} profile. This may be due to imperfect masking of the two central sources (a point-like source and a background overlapping galaxy) and an inaccurate centering of galaxy profile. Conversely, at $R > 15$~arcsec, their profile becomes noisier and decreases more slowly than our reference. Shallower profiles are expected when using circular apertures; we demonstrate this by extracting our profile using circular annuli and fixed binning.  We also show the profile derived from elliptical annuli with a fixed 2-pixel binning, which is in good agreement with our reference profile. Notably, our profile is significantly smoother than that of \citet{Muller2017}, likely due to the greater depth of the VST-SMASH data and a more precise modeling of the dwarf's geometry during the surface brightness profile determination.

For completeness, we also compare the structural parameters. They report a central surface brightness ($\mu_{\rm r,0}$) of $23.93 \pm 0.49$ mag/arcsec$^2$ compared to our value of $24.3\pm 0.1$ mag/arcsec$^2$ (after removing the extinction correction factor), an effective radius (\Rer) of 12.4 arcsec versus our geometric value of $12.7 \pm 0.4$ arcsec, and a Sérsic index ($n_{\rm r}$) of $0.93 \pm 0.25$ versus our $0.78 \pm 0.04$. The extinction-corrected magnitudes are 17.7 mag and 17 mag in $g$ and $r$ bands, which considering their uncertaintity of 0.3 mag, are consistent with our Sérsic total magnitudes of 18.04 mag and 17.49 mag. Their total extinction-corrected $g-r$ color is 0.691 mag, while we measure 0.55 mag using Sérsic magnitudes and 0.61 mag within the \Re\ in the $r$ band. Given their noisier profile, the uncertainties make their estimates consistent with ours; however, at face value there are discrepancies, likely driven by the limitations in their SB profile, as discussed above.

\subsection{Distribution of physical parameters}\label{subsec:phypar}

The distribution of the Sérsic parameters of the sample is presented in the first row of \Fig\ref{fig:histograms}, where we limit to those measured in the $r$ band and with a relative error on $R_{\rm e}$ and $n$ less than 50\%. The structural parameters in the three bands for the whole sample are reported in \Tab\ref{tab:tab2}.

The $\mu_0$ has a median value of 24.1 mag arcsec$^{-2}$, with the 16th and 84th percentiles of 22.9 and 25 mag arcsec$^{-2}$, while the effective surface brightness, $\mu_{\rm e}$, has a median value of 25.3 mag arcsec$^{-2}$, with the 16th and 84th percentiles of 24.5 and 26.2 mag arcsec$^{-2}$. Moreover, by calculating the average effective surface brightness, $\langle\mu_{\rm e}\rangle$, we obtain a median value of 24.8 mag arcsec$^{-2}$, with the 16th and 84th percentiles of 23.9 and 25.6 mag arcsec$^{-2}$, and 5th and 95th percentiles of 23.5 and 26.1 mag arcsec$^{-2}$. These values are typical of LSB galaxies and, for instance, consistent with the bulk of dwarf candidates discovered, for example, in the Fornax cluster (at a distance of $\sim$ 20 Mpc; e.g. \citealt{Venhola2019FDSVI}). Considering the small differences between the filter functions, and that the $r$- and $i$-band SB differ by only $\sim$ 0.1 mag arcsec$^{-2}$, our measurements largely overlap with the range of average SB of the $\sim 2700$ dwarf galaxy candidates found in the \emph{Euclid} Q1 data ($I_{\rm E} =  23 - 25$ mag arcsec$^{-2}$; \citealt{Marleau2025Q1Census}), while extending to somewhat fainter magnitudes. 

\begin{figure*}
\centering
\includegraphics[width=0.9\linewidth]{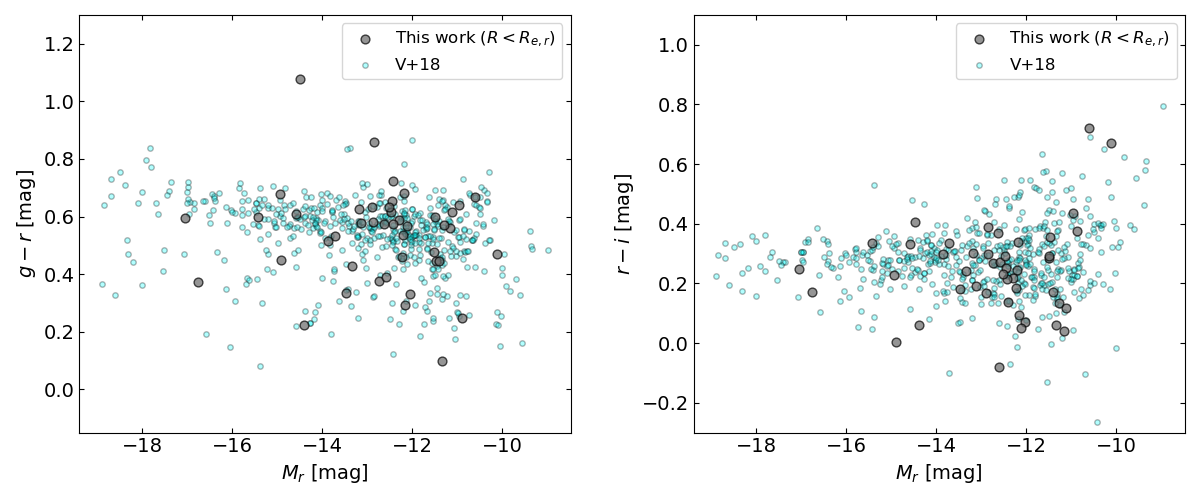}
\caption{Color magnitude diagrams. The $g-r$ and $r-i$ colors (measured within 1 \Re) are shown on the left and on the right panels, respectively, as a function of the $r$-band absolute magnitude. Grey circles with black edges  represent our sample adopting a reference distance of 26 Mpc, while cyan circles represent the dwarf galaxies from \citet[][V+18]{Venhola2018FDSIV}. According to the selection shown in \Fig\ref{fig:histograms}, dw1318-2206 is not included in the plots.}
\label{fig:CMDs}
\end{figure*}

The \Re\ shows a median value of 5.4 arcsec, with the 16th and 84th percentiles of 2.3 and 8.0 arcsec, respectively, reflecting a broad range of galaxy sizes, which can be driven by a distribution of galaxies with different intrinsic sizes and luminosities and/or different distances.\footnote{We prefer to show effective radii in arcseconds here, avoiding any conversion to physical scales since the distances of the dwarfs are unknown. In \Sec\ref{subsec:colour_and_size_lum_rel}, we will adopt reference distances for these galaxies and explicitly assess how different distance assumptions affect their physical sizes, performing comparisons with independent datasets.} The $n_r$ has a median value of 0.8, with the 16th and 84th percentiles of 0.6 and 1.1, confirming that most galaxies in the sample have relatively shallow light profiles, typical of dwarf and low-mass galaxies \citep{Caldwell83,Caon+93,GG03,Venhola2018FDSIV, Venhola2019FDSVI}. The total apparent Sérsic magnitude in the $r$ band, $\rm mag_{\rm r}$, has a median value of 19.6 mag, with the 16th and 84th percentiles of 17.8 mag and 20.7 mag, respectively, indicating a sample spanning roughly three magnitudes.

Inference of the absolute magnitudes and stellar masses (\mst) requires knowledge of the distances to the dwarf galaxies. Unfortunately, such information is unavailable for the vast majority of our sample. Therefore, to compute the absolute magnitudes and M$_\star$, we need to assume a distance. To gain some insight into the possible distances, we examine potential host galaxies within the FoV, using the HECATE sample. In our FoV, we find 33 galaxies in the HECATE catalog and therefore with measured distances. Limiting the sample to distances $\leq$100 Mpc—where we expect our dwarfs to reside in—we obtain a median distance of approximately 26 Mpc. Accordingly, we adopt this value, and discuss later the impact of varying this assumption.

The absolute $r$-band magnitude, $M_{\rm r}$, is derived from the Sérsic $r$-band apparent magnitude. The resulting distribution spans $-14.3$ mag to $-11.4$ mag (16th–84th percentiles), with a median of $-12.5$ mag, fully consistent with the dwarf-galaxy regime. \mst\ are estimated using the correlation between $g-r$ color and the stellar mass-to-light ratio (\mst/L) from \cite{Zibetti+09}, assuming a \cite{Chabrier03} initial mass function (IMF). We adopt the $g-r$ color measured within the \Re\ in the $r$ band and multiply the corresponding \mst/L by the total Sérsic luminosity to obtain the \mst. The $\log_{10} \mst/\Msun$ has a median value of 7.3, with the 16th and 84th percentiles of 6.7 and 7.9, respectively, confirming that the sample is composed of low-mass systems. These \mst\ and absolute magnitudes are in agreement with \cite{Venhola2019FDSVI}, but are smaller than those of the spectroscopic dwarf galaxy sample from \emph{Euclid}, which has a median value of $\log_{10} \mst \sim 8.6$; this difference among the two samples is mainly due to the fact that we assume a (reasonable) distance of 26 Mpc, which is significantly smaller than the distances of the \emph{Euclid} Q1 spectroscopic sample \citep{Marleau2025Q1Census}. If all distances are changed to that of NGC 5068 (i.e. 5.2 Mpc) and to a larger value of 40 Mpc, the mean absolute magnitude would change by $+3.49$ and $-0.94$ mag, respectively. These correspond to shifts in stellar mass of $-1.40$ dex and $+0.37$ dex.

Finally, to complete this first overview of the dwarf galaxy candidates properties, an important characterisation is provided by their colours, which offer a first-order indication and validation of their nature. In the last two panels of \Fig\ref{fig:histograms}, we show the distributions of $g-r$ and $r-i$ colors, measured both within a fixed radius of 5 arcsec and within the \Re\ in the $r$ band. The two sets of colours are broadly consistent, although the measurements within \Re\ produce a narrower distribution. In particular, the median $g-r$ and $r-i$ colours within \Re\ are 0.57 mag and 0.24 mag, with 16th–84th percentiles of 0.37–0.64 mag and 0.10–0.36 mag, respectively. As clearly seen in the histograms, the colors are coherent with the typical color ranges of dwarf candidates in the Fornax cluster \citep{Venhola2018FDSIV,Venhola2019FDSVI}.

\subsection{Colour- and size-luminosity relations}\label{subsec:colour_and_size_lum_rel}

To provide a more physically motivated characterisation of our galaxy sample, in this section we analyze two sets of scaling relations: the colour–magnitude relation, and the correlation between structural parameters and magnitude.

In \Fig\ref{fig:CMDs}, we show the colour–magnitude diagrams, plotting $g-r$ and $r-i$ colors (measured within 1 \Re) as a function of the $r$-band absolute magnitude. This allows a more detailed comparison with the sample of dwarf galaxy candidates in the Fornax cluster \citep{Venhola2018FDSIV,Venhola2019FDSVI}, for which colours are measured within the same aperture. Consistent with the results in \Sec\ref{subsec:phypar}, we find discrete (also if not perfect) agreement with the Fornax dwarf galaxy candidates (see more details in \App\ref{app:comparisons_statistics}). While the Fornax sample benefits from larger statistics, particularly at the bright end ($M_{\rm r} < -14$), our smaller sample is consistent with this independent dataset, confirming the reliability of our sample selection.

\begin{figure*}
\centering
\includegraphics[width=0.99\linewidth]{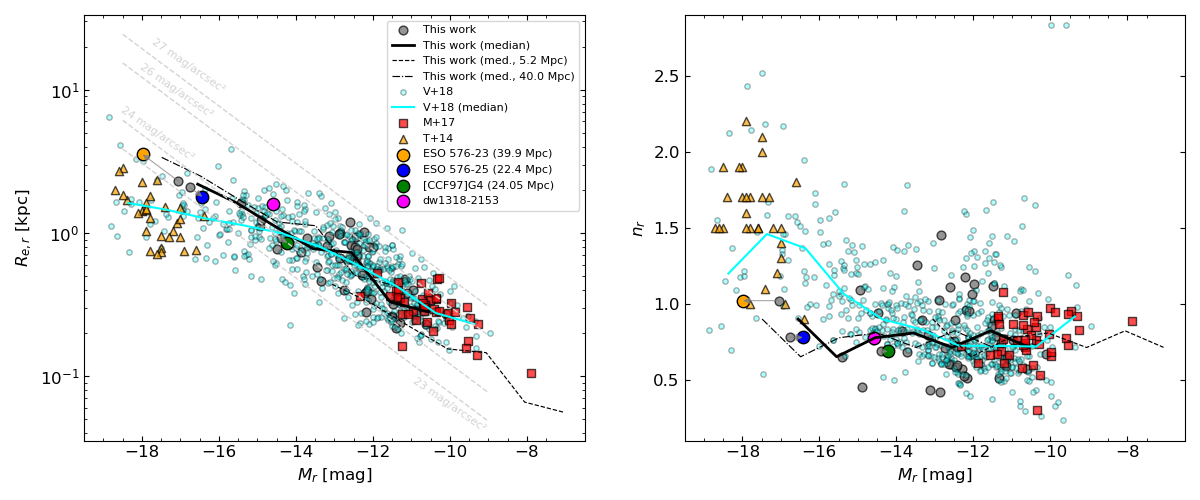}
\caption{The \Re\ (left panel) and Sérsic index (right panel) in the $r$ band  as a function of the $r$-band absolute magnitude. Grey circles with black edges represent our sample adopting a reference distance of 26 Mpc. Black solid lines indicate the median trends. Dashed and dot-dashed black lines show the effect of assuming distances of 5.2 and 40 Mpc, respectively. Cyan circles correspond to dwarf galaxies from \citet[][V+18]{Venhola2018FDSIV}, with the cyan lines tracing the corresponding median trends, red squares are for dwarfs in \citet[][M+17]{Muller2017}, and orange triangles are for dwarf ellipticals from \citet[][T+14]{Toloba+14_II}. Data points for galaxies known in the literature—dw1317-2137 (ESO 576-23), dw1318-2041 (ESO 576-25), dw1320-2200 ([CCF97]G4), and dw1318-2153—are also plotted as colored points. In particular, for the two galaxies with measured distances, the \Re\ and absolute magnitude have been recomputed, with an arrow connecting the value at the reference distance to the value at the measured distance from the HECATE catalog. Dashed grey lines in the left plot indicate tracks of constant effective SB. According to the selection shown in \Fig\ref{fig:histograms}, dw1318-2206 is not included in the plots.}
\label{fig:Re_n_vs_Mr}
\end{figure*}

Moreover, a size–luminosity or size–mass relation is known to exist across a wide range of rest-frame luminosities and \mst, and it is also evident in dwarf galaxies \citep[e.g.,][]{Papaderos+96,Schombert+06,Amorin+09,Venhola2019FDSVI,LaMarca+22_Hydra-I}. In the left panel of \Fig\ref{fig:Re_n_vs_Mr}, we show the \Re\ in the $r$ band versus the absolute magnitude, along with the median trends, which confirm that the brightest (and most massive) dwarfs galaxy candidates are also larger in size. For comparison, lines corresponding to different values of the $\mu_{\rm e}$ are overplotted, with our objects largely falling within the track defined by 23–27 mag arcsec$^{-2}$. Our sample here has properties that are similar to the dwarf galaxies in Fornax \citep{Venhola2018FDSIV,Venhola2019FDSVI}. In the right panel of \Fig\ref{fig:Re_n_vs_Mr}, we compare the Sérsic index and absolute magnitude. Our sample does not exhibit any significant trend, likely due to limited statistics, and presents $n < 2$ at all magnitudes. The comparison with \cite{Venhola2019FDSVI} dwarfs is reasonable, except for the larger scatter in our sample and the trend at high luminosity ($M_{\rm r} < -14$), where the \cite{Venhola2019FDSVI} sample shows a rising $n$.\footnote{Applying the Kolmogorov-Smirnov test to \Re\ (adopting the reference distance) and $n$, we find significant agreement, with $p > 0.05$.} 

Given that the distances to our galaxies are unknown and the adopted value is only an assumption, \Fig\ref{fig:Re_n_vs_Mr} also shows the impact of changing the distance to 5.2 Mpc (the distance of NGC~5068, the nearest galaxy in this field) and 40 Mpc, a representative distance for other galaxies in the field. To avoid cluttering the plot, we show only the median trends for these two distances. Overall, while a distance of 40 Mpc produces a physically plausible size–mass relation, consistent with the Fornax dwarfs, we could largely exclude that most galaxies are closer and physically associated with NGC~5068. The similarity between our structural parameters and those in \cite{Venhola2018FDSIV} is also investigated in \App\ref{app:comparisons_statistics} using statistical tests.

Figure~\ref{fig:Re_n_vs_Mr} also highlights the four previously known galaxies. For the two with distances available in the HECATE catalog, we indicate the shift of their positions in both panels of \Fig\ref{fig:Re_n_vs_Mr}. In terms of \mst, ESO 576-23 (dw1317-2137) has $\log_{10} \mst/\Msun = 9.15$ at the reference distance, which increases to 9.5 when adopting the HECATE distance of 39.9 Mpc. Conversely, ESO 576-25 (dw1318-2041) has a reference mass of 8.6, which slightly decreases to 8.5 using the HECATE distance.

We also compared our results with structural parameters of bright Virgo dwarf ellipticals from \cite{Toloba+14_II} and dwarf candidates in the Centaurus group from \cite{Muller2017}. To convert the observed \Re\ into physical scales, we adopted distances of 15.85 Mpc for the former and 4.5 Mpc for the latter. The Virgo cluster sample occupy the high-mass end of the dwarf distribution from \cite{Venhola2019FDSVI}, and only slightly overlapping with the brightest dwarf galaxy candidates in our sample. The overlap would increase if our galaxies were, on average, located at larger distances. In terms of size and n, these massive dwarfs follow the trends reported by \cite{Venhola2019FDSVI}.

We also considered the results of \cite{Muller2017}, assuming the distance of Cen A, with their systems populating the low-mass end of both our galaxy sample and that of \cite{Venhola2019FDSVI}. We would better overlap with them if smaller distances were adopted in our case (dashed black line). A more significant discrepancy emerges in the $n$ values: although there is some overlap with the higher values in our sample, their dwarfs are systematically found with n $\geq$1, and no galaxies with lower indices are present in their study.

\begin{figure*}
\centering
\includegraphics[width=0.9\linewidth]{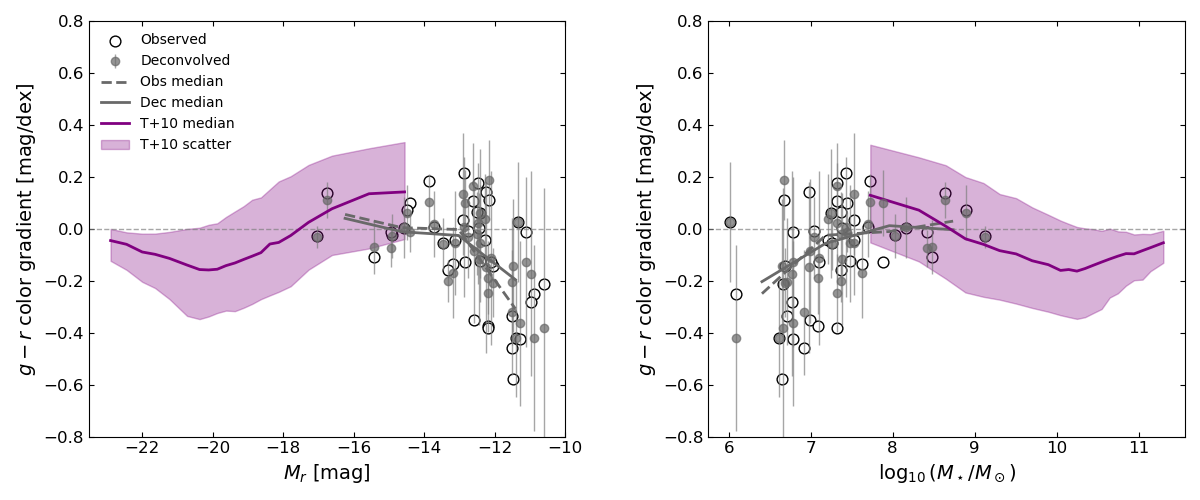}
\caption{$g-r$ colour gradients as a function of absolute magnitude (left panel) and stellar mass (right panel). Gradients measured from the observed data (open circles) and from the deconvolved Sérsic fits (gray circles with black edges) are shown, together with their median trends (dashed and solid lines, respectively). Uncertainties on the deconvolved Sérsic fits are also plotted. Colour gradients from \cite{Tortora+10CG}, based on the Sérsic fits in the \cite{Blanton+05_NYU} dataset, are indicated by the purple line, with the shaded region representing the 16–84th percentile range. Only the 44 candidates with available measurements of both $M_{\rm r}$ and \mst\ and with the smallest uncertainties in the structural parameters are shown (see text).} 
\label{fig:gradients}
\end{figure*}

\subsection{Colour gradients}\label{subsec:colour_gradients}

Colour and stellar population gradients, in addition to integrated quantities, can provide further insights into galaxy evolution, as they trace the stellar population variations from the center to the outskirts, allowing us to disentangle different processes that affect stellar populations differently as a function of galacto-centric distance \citep[see][and reference therein]{Tortora+10CG}.

Colour gradients can be parameterized through the ratios of effective radii and Sérsic indices measured in two different bands, as colour gradients naturally induce wavelength-dependent variations in these parameters \citep[][]{Vulcani+14,Baes+24_II,Quilley+25_Euclid_Q1}.

Here, however, to combine the information from both \Re\ and $n$, we instead measure the colour gradients directly from the colour profiles in the central regions, following \citet{Tortora+10CG}. Using the $g-r$ and $g-i$ profiles, we calculate the gradient as the slope ($b$) in the relation: $\rm colour = a + b \log(R / \Rer)$, fitted to the data. Gradients are derived from both the observed data and the deconvolved Sérsic fits. The latter mitigates seeing effects, which are particularly relevant for the smallest galaxies in the sample.

We focus on the $g-r$ colour, which is constructed using the deepest bands in our dataset, but the conclusions reached using $g-i$ are very similar, although with a slightly larger scatter. The results, shown in \Fig\ref{fig:gradients}, are plotted against both absolute magnitude and \mst. In this section, we limit the analysis to the 42 dwarf candidates with available measurements of both $M_{\rm r}$ and \mst, and with relative errors smaller than 50\% in the $g$- and $r$-band $R_{\rm e}$ and $n$ measurements, in order to exclude highly uncertain colour gradients. To place our findings in a broader context, we compare them with those of \citet{Tortora+10CG}, who used Sérsic fits for $\sim 50\,000$ local galaxies from the NYU Value-Added Galaxy Catalog (NYU-VAGC; \citealt{Blanton+05_NYU}). This reference sample spans a wide range of galaxy luminosities and \mst, reaching down to $M_{\rm r} \sim -13$ and $\log_{10} \mst/\Msun \sim 7$.

For the general bright population in \cite{Tortora+10CG}, composed of massive ellipticals and spirals, gradients are negative (redder in the center), while they turn positive for the faintest/least massive galaxies (bluer in the center). Notably, the colour gradients show a minimum at $M_{\rm r} \sim -20$ and $\log_{10} \mst/\Msun \sim 10.3$. These trends are in good agreement with the analysis by \cite{Liao_Cooper2023}, based on DESI data. The mass at which the minimum occurs corresponds to the “golden mass”, a characteristic scale in galaxy evolution associated with the peak of star formation (SF) efficiency and extrema in several scaling relations (see \citealt{Tortora+25_CASCOIII}, and references therein).

For our dwarf candidates the colour gradients exhibit a wide scatter. However, we find that, particularly for the deconvolved data, the gradients are on average negative, with steeper negative slopes at lower \mst. In particular, at masses $\log \mst/\Msun < 7.5$, we find a deconvolved median $g-r$ ($g-i$) gradient of $-0.12$ ($-0.13$) $\rm mag/dex$, with a $1\sigma$ scatter of $0.16$ ($0.2$) $\rm mag/dex$, and corresponding weighted means of $-0.09 \pm 0.03$ ($-0.15 \pm 0.04$) $\rm mag/dex$. Mostly null/positive values are found instead in the most massive dwarfs ($\mst \gsim 10^{8}\, \rm \Msun$), as indicated by the median trends and consistent with the results of \cite{Tortora+10CG}. This latter behaviour for the most massive galaxies in our sample is consistent with the gradients found in a sample of 211 nearby dwarf galaxies with masses in the range $10^{8} < \mst < 10^{9.5}\, \rm \Msun$ in low-density environments, where inner gradients range from nearly zero values for late-type dwarfs to positive values ($\sim 0.25 \, \rm mag/dex$) for early-type dwarfs \citep{Lazar+24_structure_dwarfs}.

Different processes can shape dwarf galaxy evolution. Early monolithic collapse models predict strong central metallicity gradients, while supernova feedback can efficiently quench the SF and expel gas, determining similar gradients \citep[e.g.,][]{Chiosi_Carraro2002}. Environmental effects, such as ram-pressure stripping and galaxy harassment \citep[e.g.,][]{Mayer+01}, would have an effect on the morphology and nature of the dwarfs, but would primarily affect the outer regions ($> \Re$), and play a minor role in the central regions, where the potential well is deeper \citep{Boselli+08}.

Since our analysis focuses on central regions, the negative colour gradients observed would be mostly driven by negative metallicity gradients, produced by a combination of early collapse and supernova feedback \citep{Tortora+10CG}. Age gradients in dwarf galaxies are generally weak or mildly positive, with younger populations preferentially located toward the centre, while metallicity gradients are typically negative, and there could also be some contribution from dust extinction in the centres \citep[e.g.,][]{Koleva+11_gradients,deBoer+11_Sculptor,Annibali_Tosi2022}, thus, negative colour gradients are most naturally attributed to metallicity variations. This connection is also consistent with results in \cite{Mercado+21}, who comparing FIRE-2 cosmological
baryonic zoom-in simulations of isolated galaxies with LG dwarf galaxies, have shown that
these objects present stellar metallicity gradients which are driven by the baryon/feedback cycle rather than by 
environmental effects. In particular their gradients are shaped by two competing processes: (a) the gradual ‘puffing up’ of old, metal-poor stars due to feedback-driven potential fluctuations, and (b) the late-time accretion of extended, metal-rich gas, which fuels the formation of metal-rich stars. The positive colour gradients found in some galaxies could be related to dominant younger stellar populations in the center.

The environment may play an important role in shaping the morphology of our predominantly smooth systems. Environmental effects can suppress SF in these galaxies, potentially transforming star-forming dwarfs (irregulars) into more quiescent systems, such as dwarf ellipticals and dwarf spheroidals \citep{Boselli+08}. Although limitations in resolution and in the visual selection process can affect classification in some cases, and we deliberately exclude smaller and fainter galaxies, which can be highly contaminated by background objects, this scenario appears consistent with our findings: we observe a larger fraction of smooth galaxies compared to more structured ones, with approximately 75\% of our sample exhibiting a smooth morphology. These galaxies can be likely located in the outskirts of the Virgo Supercluster, an intermediate environment between the dense centers of the Perseus cluster \citep{Marleau2025PerseusERO}, where dwarf ellipticals dominate, and the more generic fields of the \emph{Euclid} Q1 survey \citep{Marleau2025Q1Census}, where dwarf ellipticals constitute about 58\% of the population, with higher fractions near potential host galaxies, or the field environment in \cite{Lazar+24}, where the fraction of early-type dwarfs and featureless dwarfs is $\sim 53\%$. The smooth nature of most of our galaxies may also classify some of them as dwarf spheroidals, implying lower masses and luminosities and supporting a possible association with NGC~5068. The connection with possible hosts and the role of the environment will be further discussed in the next subsection.

Regarding colour gradients in the external regions, we focus on the $g-r$ colour only, since $g-i$ is less reliable due to the lower S/N in the $i$-band and the higher sky contamination at large radii. Computing the gradients out to $5\,R_{\rm e,r}$, we find that for the deconvolved profiles the median value is $0.1$, with a large scatter of $0.7$. However, when accounting for uncertainties, we obtain a weighted mean of $-0.1 \pm 0.2\, \rm mag/dex$. This suggests a large intrinsic scatter, potentially indicating a different origin and environmental dependence, but with a mild preference for negative colour gradients.

\begin{figure*}
\centering
\includegraphics[width=0.9\linewidth]{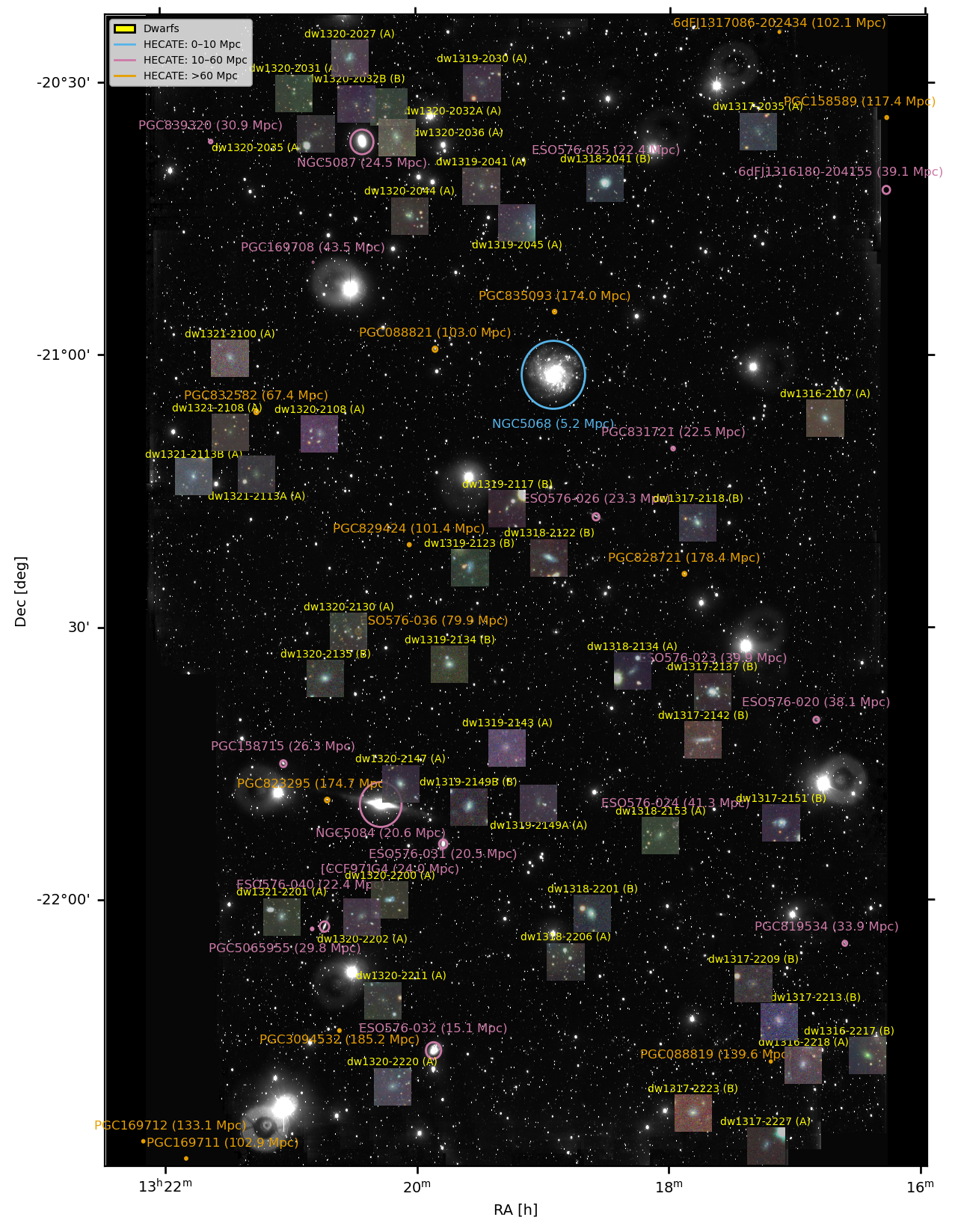}
\caption{Spatial distribution of dwarf galaxy candidates and their potential host galaxies, superimposed to the $r$-band VST-SMASH mosaic. In the VST–SMASH background image, the brightest galaxies are clearly visible, along with saturated stars and their reflection halos, which are typical artefacts in VST images. Coloured circles indicate the positions of possible host galaxies from the HECATE catalog, with radii set to the geometric mean of the major and minor axis radii reported in the catalog, and colored according to their distance (as shown in the legend). The RGB cutouts of the dwarf candidates, located at their positions, are overplotted at arbitrary sizes for visualization purposes, with their assigned grades indicated. North is up, and east is to the left.}
\label{fig:spatial_distribution}
\end{figure*}

\begin{figure*}
\centering
\includegraphics[width=0.9\linewidth]{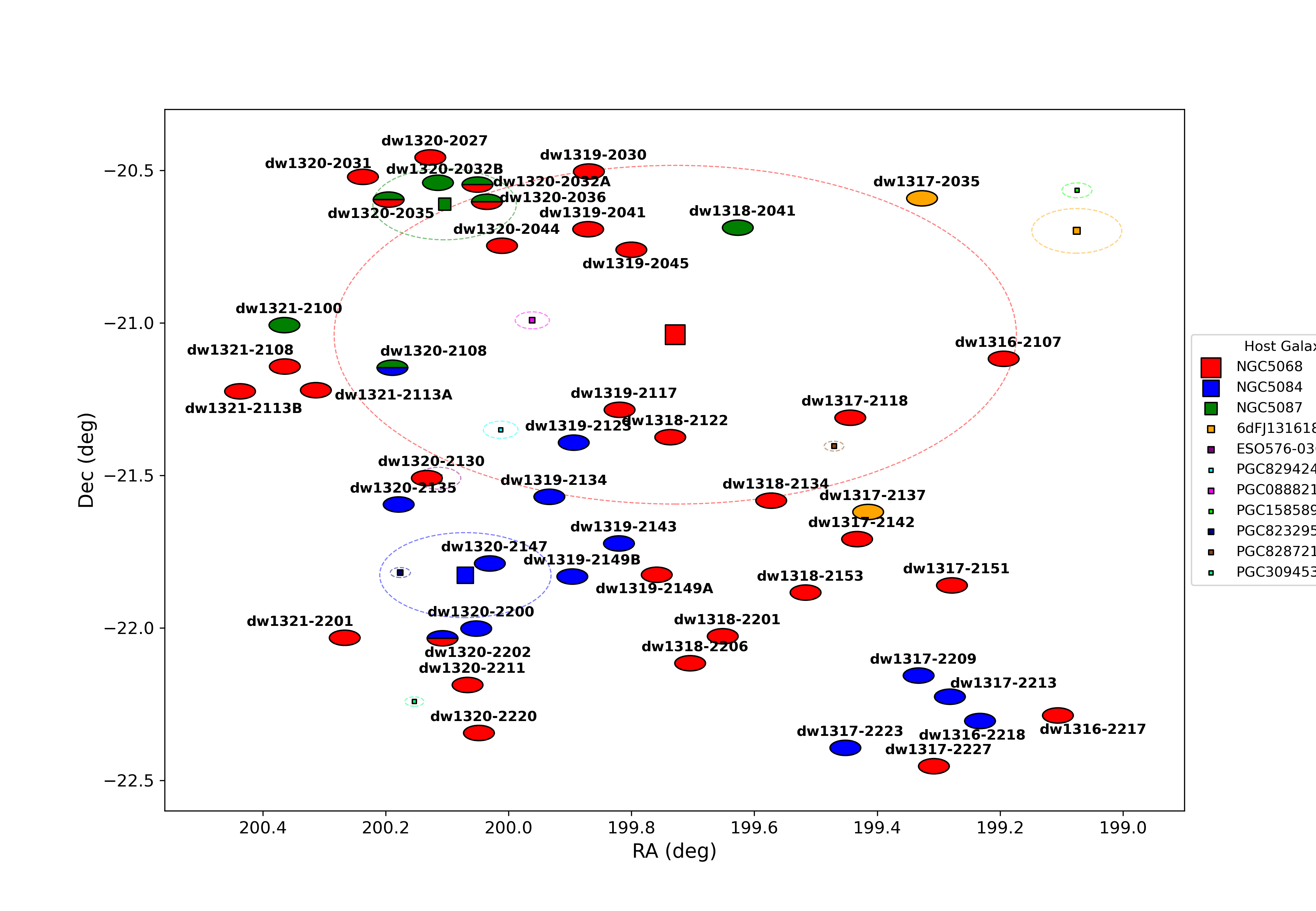}
\caption{Spatial distribution of dwarf galaxy candidates (ellipses) and their associated host galaxies (squares). Each dwarf candidate is coloured according to the host with the highest membership probability. If multiple hosts have membership probabilities within 25\% of the highest probability, the dwarf galaxy shape is divided into wedges coloured according to each host. The dashed ellipses correspond to a radius of 50 kpc.} 
\label{fig:association}
\end{figure*}

\subsection{Spatial distribution}

The morphology of dwarf galaxies appears to be strongly influenced by their environments: early-type dwarfs are preferentially found in groups and clusters, where environmental processes such as ram-pressure stripping, tidal harassment, and starvation play a key role, while dwarf irregulars are more evenly distributed across the field \citep[see, e.g.,][and references therein]{Marleau2025PerseusERO,Marleau2025Q1Census}. Mapping the spatial distribution of the identified dwarf galaxy candidates can allow us to assess whether they are associated with specific host galaxies, and therefore to evaluate the role of environmental processes in driving their structural and morphological evolution. 

Given the area of the mosaic, approximately 2.7 \sqd, we find about 15.6 dwarf candidates per \sqd\ (8.9 and 6.7 candidates per \sqd\ with A and B grade, respectively). This number density is lower than that found by \emph{Euclid}, which identified 188 dwarf candidates per~\sqd\ in generic environments \citep{Marleau2025Q1Census}, and about $1100$ dwarfs within the $0.7~\sqd$ field of view at the centre of the Perseus cluster. These differences are clearly due to a combination of factors: the higher Euclid resolution and in particular the fact that both samples (in particular the former) are at larger distances, naturally yielding a larger number of dwarfs per unit area. Moreover, while our environments are broadly consistent with the \cite{Marleau2025Q1Census} sample, in a dense environment such as the Perseus cluster we naturally expect a higher number of dwarf galaxies. Our values are comparable to the $\sim 22$ dwarf candidates per \sqd\ reported for the Fornax cluster \citep{Venhola2018FDSIV,Venhola2019FDSVI}. Since this latter analysis was carried out in the center of the Fornax cluster, where a larger number of dwarfs is expected, our findings highlight the high efficiency of VST-SMASH in identifying dwarf candidates. In comparison with other large-scale searches, our number density is much higher than the 0.7 dwarf elliptical candidates per \sqd\ found by \cite{Paudel+23} observed over the 7643 \sqd\ of the Legacy Survey. Similarly, \cite{Muller2015} inspected 60 \sqd\ around the M83 subgroup and discovered 16 new dwarf galaxies, corresponding to 0.27 candidates per \sqd. Extending the survey by another 500~\sqd\ in the same subgroup, \cite{Muller2017} found 41 additional dwarf candidates, corresponding to 0.082~per~\sqd\ over the 500~\sqd\ area.

However, it is even more important to understand the distribution of the candidates relative to the possible hosts, i.e., the $\sim 30$ galaxies in our FoV. To provide a complete view of the dwarf distribution, we highlight the positions of both the host galaxies (coloured by their distance) and the dwarf candidates (in yellow) on the $r$-band mosaic shown in \Fig\ref{fig:spatial_distribution}.

Given that the distance of the dwarf candidates is unknown, we cannot unambiguously determine their association with a specific galaxy, nor whether they are truly isolated. However, we first make a preliminary assessment of galaxy membership by analyzing the spatial distribution of the dwarf candidates, using the positions of the possible host galaxies, and then proceed with a more quantitative analysis including additional observables.

When examining their projected separation, we observe clear overdensities around the two largest Virgo Supercluster members in the FoV, namely NGC~5084 (south-east) and NGC~5087 (north-east). In the vicinity of NGC~5087, three additional cluster members are found --- ESO~576-032, ESO~576-036, and ESO~576-040 --- which, together with their larger companion, could plausibly act as the main hosts of dwarfs in this area. Two further overdensities of dwarf candidates are identified in the central-east and south-west regions of the FoV, with nearby potential hosts located at larger distances than the previous members. 
NGC~5068 represents a rather different case: although the average size--luminosity relation of the dwarf candidates, when placed at its distance of 5.2~Mpc, is not consistent with literature expectations (unless we assume that these objects are intrinsically fainter than those in \citealt{Venhola2019FDSVI}), their spatial distribution suggests that many of these systems could, in fact, be satellites of this galaxy. 

To perform a more quantitative analysis, we developed an empirical procedure that associates each dwarf with its most probable host. We define a probability $\mathcal{P}$ as: 
\begin{equation}
    \mathcal{P} = \mathcal{P}_{\rm sep} \times \mathcal{P}_{\rm size} \times \mathcal{P}_{\rm mag} \times \mathcal{P}_{\rm dist-known}.
\end{equation}
Each term is defined as follows:
\begin{itemize}
\item $\mathcal{P}_{\rm sep}$ is based on the projected separation of the dwarf galaxy from the $i$-th host candidate, $d_{\rm sep}$. We adopt as $\mathcal{P}_{\rm sep}$ the projected number-density profile of satellite galaxies around Luminous Red Galaxies derived by \cite{Tal+12}, normalized such that its maximum value is equal to 1. This function decreases steeply toward zero, reaching values below 0.1 at distances larger than $\sim 60$ kpc and asymptotically approaching zero at larger radii. Although this function does not account for possible variations of the satellite number-density profile with galaxy type and mass, it is qualitatively consistent with what is observed in the Milky Way, where most satellites are concentrated within $\sim 50$ kpc \citep{mcconnachie12}. Future implementations may further refine this term by adopting host-dependent satellite distributions that account for host galaxy mass and morphology. 
\item $\mathcal{P}_{\rm size}$ refers to the probability associated with the dwarf galaxy size, parametrized by the fitted $R_{\rm e}$. To construct this probability function, we combine two data samples which includes both dwarf spheroidals and a more generic dwarf population in clusters from: (a) the Local Volume Database (LVDB; \citealt{pace2024localvolumedatabaselibrary}) and (b) the Extended Virgo Cluster Catalog (EVCC; \citealt{Kim+14_EVCC}). We use the half-light radii to build a histogram of the size distribution and transform it into a probability distribution by normalizing its maximum value to 1. The distribution of half-light radii is strongly peaked at small values and drops below 0.1 for radii larger than $\sim 2.3$ kpc.
\item $\mathcal{P}_{\rm mag}$ concerns the comparison with the distribution of absolute magnitudes from \citet{Venhola2019FDSVI}. We adopt a simple criterion: the probability is set to zero when the dwarf galaxy absolute magnitude, computed at the assumed host distance, falls outside the absolute magnitude range covered by the relation. In particular, as a conservative choice, we adopt the range $[M_{\rm r,min} - 0.5,\; M_{\rm r,max} + 0.5]$. A more refined treatment could assign probabilities based on the dwarf galaxy offset, along the y-direction (i.e. the size in kpc), from the median relation. 
\item $\mathcal{P}_{\rm dist-known}$ is set to 1 for all galaxies without a distance measurement. For galaxies with an estimated distance, it is set to 1 if the distance difference with respect to the $i$-th host is less than 5 Mpc, and 0 otherwise.
\end{itemize}

We note that, although our approach is quantitative, the choice of the individual probability terms remains somewhat arbitrary, and in the present analysis we neglect the possibility that some galaxies may be isolated. In future work, we plan to use cosmological simulations to develop a refined and physically motivated probabilistic scheme.

The results are presented in \Fig\ref{fig:association}, where only the largest host galaxies, with a Holmberg radius $\geq$ 5 kpc, are considered in order to simplify the associations and retain only the systems most likely to act as hosts. Dwarf candidates are linked, in colour to the HECATE galaxies with the highest membership probability. Despite the inevitable arbitrariness of our probability construction, given their mass, NGC~5068, NGC~5084, and NGC~5087 looks to be the most likely hosts for the majority of dwarf candidates.            

Focusing on the edge-on lenticular galaxy NGC~5084, we identify a potential association with at least 10 dwarf candidates. Some of these appear to be located in a region roughly perpendicular to the disc, suggesting a possible analogy with the VPOS, the plane of satellites observed around the Milky Way. However, confirming this scenario would at least require radial velocity measurements of these objects, as well as a kinematic study of their relation to NGC~5084 (e.g., \citealt{Pawlowski2012}). Approximately 5 dwarf galaxies show a high-probability association with NGC~5087. Given its closer distance, NGC~5068 is the most suitable host for most of the remaining dwarf galaxies.

\section{Conclusions}\label{sec:conclusions}

Exploiting the excellent image quality of the VST telescope at Cerro Paranal,  VST-SMASH aims to explore a sample of galaxies in the nearby Universe ($<10$ Mpc) with unprecedented depth \citep{Tortora+24_VST-SMASH}. In this paper, we have analyzed one of the fields observed, containing several galaxies in the local Universe: the large face-on spiral NGC~5068 at a distance of 5.2 Mpc, and the more distant edge-on lenticular galaxy NGC~5084 together with the elliptical galaxy NGC~5087, which are located in the peripheral regions of the Virgo Supercluster. This FoV covers an area of about 2.6~deg$^2$, with an average seeing of FWHM $\sim 1$ arcsec and $1\sigma$ SB limits of 30.7, 30.2, and 29.4 $\rm mag/arcsec^{2}$ in the $g$, $r$, and $i$ bands, respectively.

We carried out a visual inspection of the field, based on a two-step procedure of initial tagging and subsequent vetting, which led to the identification of 47 dwarf galaxy candidates: 30 classified as high-probability dwarfs, and 17 as likely dwarfs. After performing accurate automatic and manual masking of the regions around the candidates, we performed surface photometry and fitted 1D Sérsic profiles. Only four of these systems had previously been reported in the literature. For one of them, dw1318-2153, we compared our structural parameters with those obtained by \citet{Muller2017}, finding good agreement.

The dwarf candidates exhibit physically plausible colours and structural parameters. Their median colours, $g-r = 0.57$ mag and $r-i = 0.24$ mag  (measured within the \Re\ in the $r$ band), are fully consistent with Fornax cluster dwarf galaxy populations \citep[e.g.][]{Venhola2019FDSVI}. Even the more scattered colours remain broadly compatible with values reported in the literature. We also examined scaling relations between  \Re, n, and absolute luminosity, assuming a reference distance of 26 Mpc. The results show good consistency with previous studies \citep{Venhola2019FDSVI}. The correlations remain valid if distances up to 40 Mpc are adopted (i.e., the most distant Virgo members), but become more incompatible if the nearest distance of 5.2 Mpc (NGC~5068) is assumed, since in that case the galaxies would appear too faint. This disfavors NGC~5068 as the host of the majority of dwarfs in the field, unless we conclude that our galaxies are intrinsically systematically fainter than those found in the Fornax cluster. We have also measured and analyzed color gradients in the central regions of the galaxies, against absolute magnitude and \mst. These gradients show a wide scatter, with both positive and (mostly) negative values, which we interpret within the broader framework of galaxy formation. 

Finally, we analyzed the spatial distribution of the dwarf candidates and their potential association with host galaxies (selected from the HECATE catalog, \citealt{Kovlakas+2021_HECATE}). Based on their projected proximity and other empirical criteria, we confirm that the largest galaxies in the field (NGC~5084, NGC~5087, and NGC~5068) are plausible hosts for many of our dwarf galaxy candidates. In particular, given its close distance, several candidates are located at physically plausible separations from NGC~5068, even though a comparison with the size–luminosity relation might suggest otherwise.

Future spectroscopic follow-up and/or higher-resolution, deeper imaging will be crucial to further constrain the nature of these candidates. Such data would allow us to characterize their stellar populations, obtain precise distance and velocity measurements, and thereby confirm their association with host galaxies. This would enable detailed studies of their spatial distributions, satellite counts, and potential alignments, and provide valuable tests of cosmological models, including predictions for the incidence of planes of satellites and the DM nature \citep[e.g.][]{Muller+18_science,Sawala+23_plane_of_satellites,Pawlowski+23_memorie}.

\begin{acknowledgements}
We thank the anonymous referee for their constructive comments, which significantly improved the presentation and the results of this paper. R.R. acknowledges financial support grants through INAF-WEAVE StePS funds and through PRIN-MIUR 2020SKSTHZ. We acknowledge the usage of the HyperLeda database (http://leda.univ-lyon1.fr). Based on data
collected with the INAF VST telescope at the ESO Paranal Observatory.
\end{acknowledgements}

\bibliography{myrefs}

\begin{appendix}

\section{Single-band images and masks}\label{app:single-band_images}

\begin{figure*}
\centering
\includegraphics[width=0.95\linewidth]{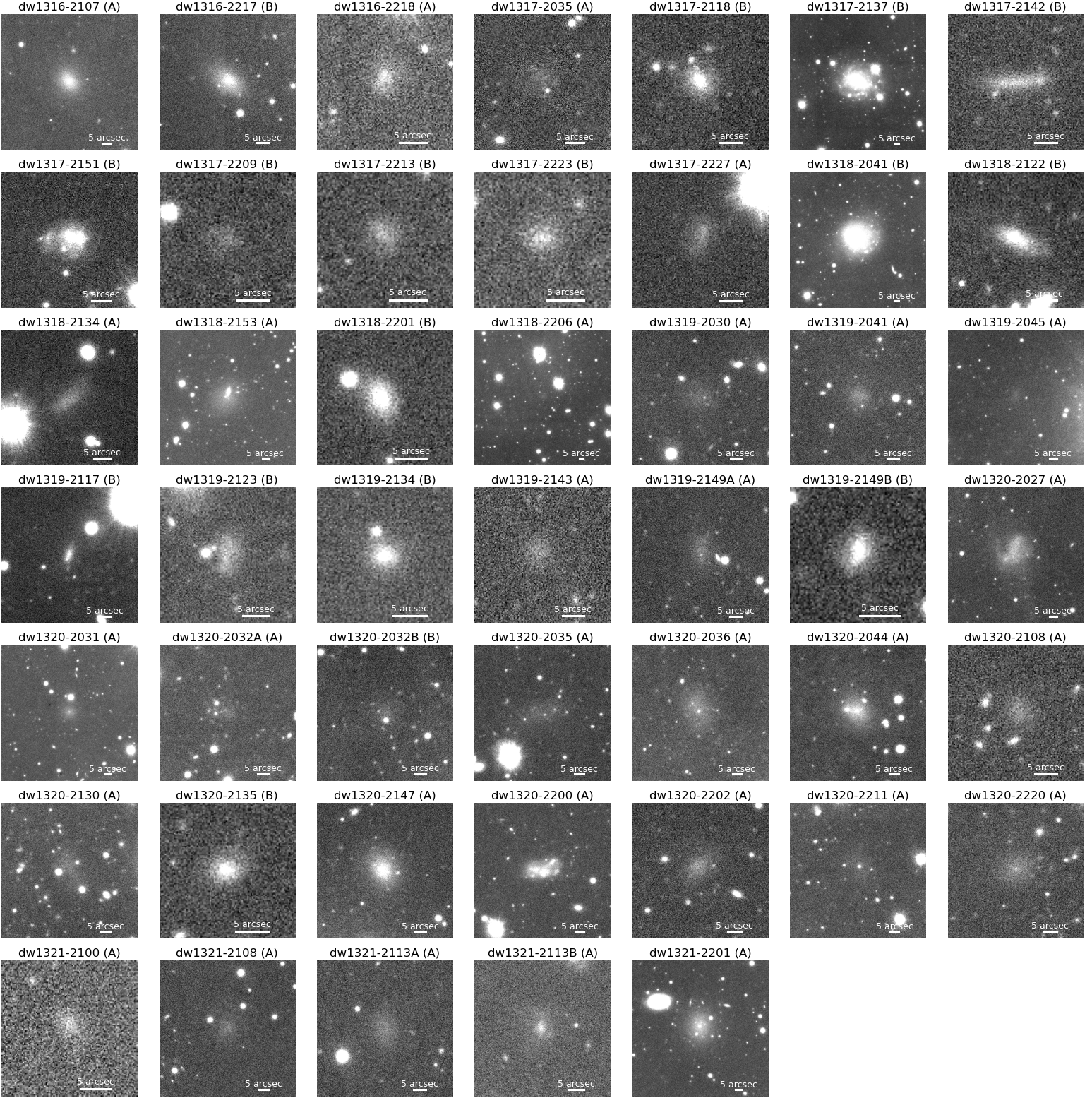}
\caption{$r$-band images of the selected dwarf candidates after the tagging and vetting stages of our visual classification. The horizontal white bar corresponds to 5 arcsec. Each cutout has a size equal to 10 times the effective radius of the galaxy in the $r$ band.
The assigned vote is given in parentheses.}
\label{fig:r-band_images}
\end{figure*}

\begin{figure*}
\centering
\includegraphics[width=0.95\linewidth]{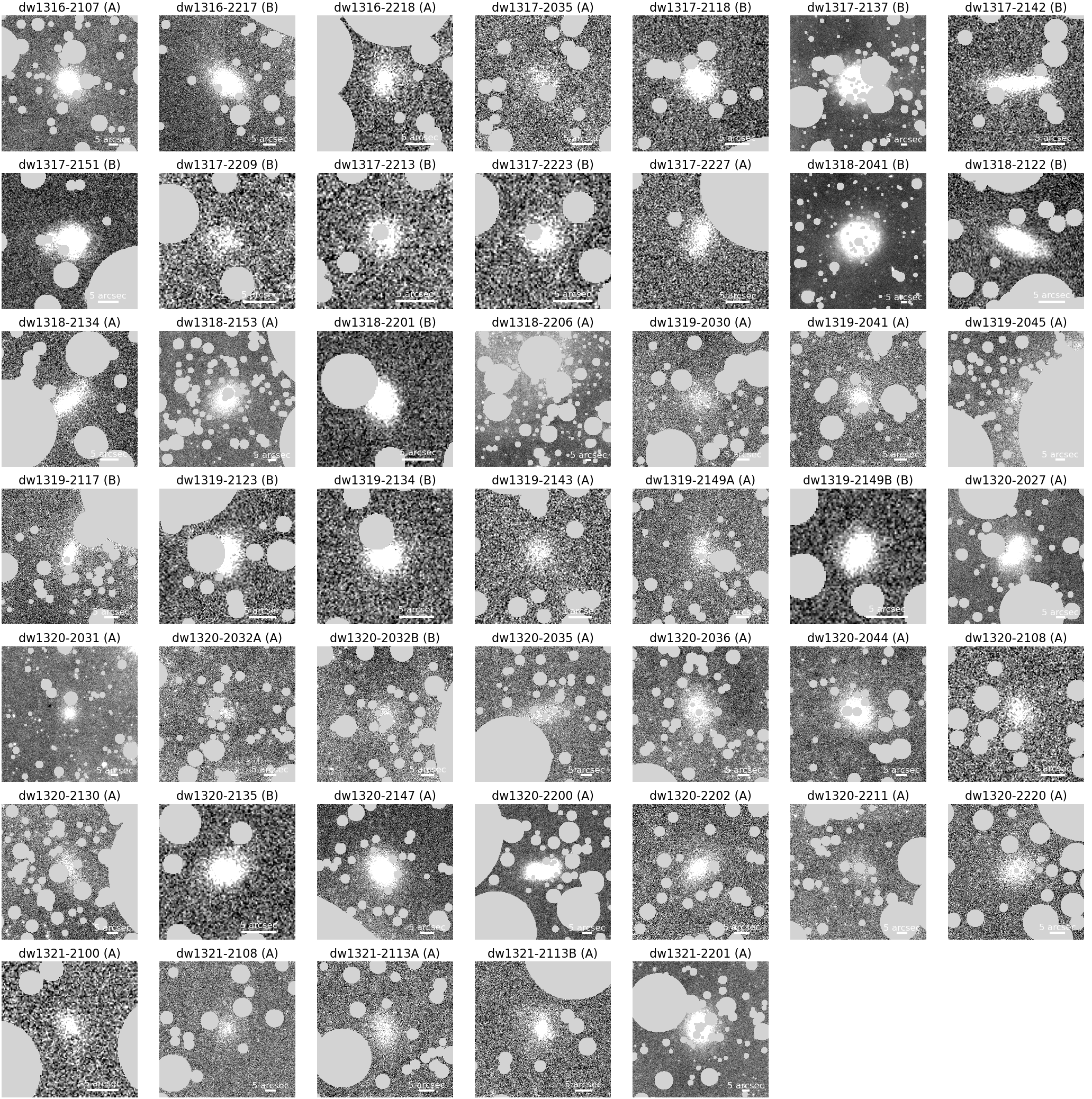}
\caption{Masked $r$-band images of the selected dwarf candidates after the tagging and vetting stages of our visual classification. The horizontal black bar corresponds to 5 arcsec. Each cutout has a size equal to 10 times the effective radius of the galaxy in the $r$ band.
The assigned vote is given in parentheses.}
\label{fig:masked_r-band_images}
\end{figure*}

$r$-band images and masks superimposed on the $r$-band images are shown in \Figs\ref{fig:r-band_images} and \ref{fig:masked_r-band_images}.

\section{SB profiles}\label{app:SB_profiles}

\begin{figure*}
\centering
\includegraphics[width=1\linewidth]{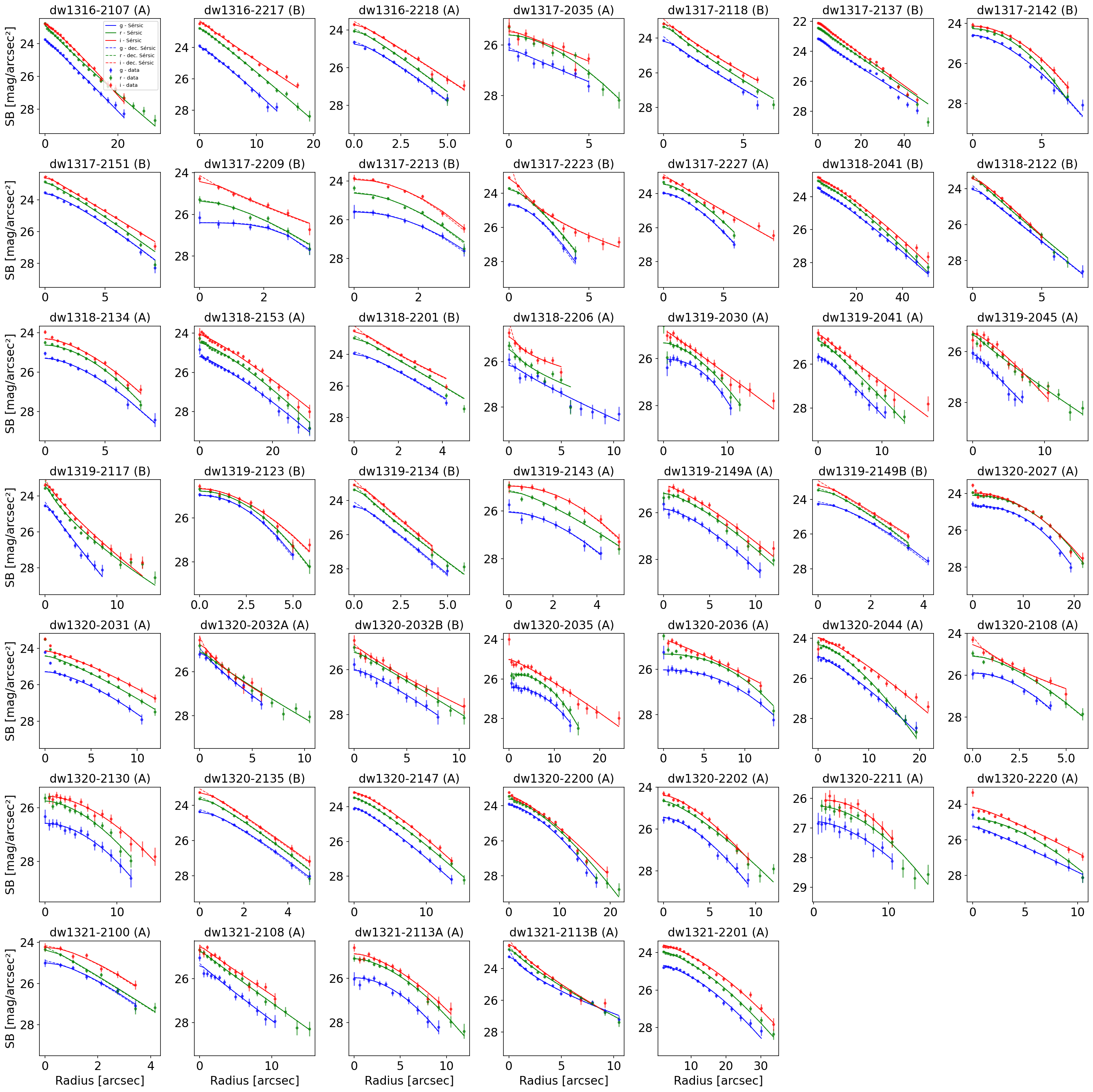}
\caption{1D SB profiles of the dwarf candidates. Data points with error bars are shown as dots, while the PSF-convolved and deconvolved Sérsic fits are plotted as solid and dashed lines, respectively. Profiles in the $g$, $r$, and $i$ bands are shown in blue, green, and red, respectively. Structural parameters are listed in Table \ref{tab:tab2}. The single-component Sérsic profile provides a reasonable approximation of the light profiles of almost all the dwarf candidates. In general, we fail to obtain a credible fit in only a very small number of cases.}
\label{fig:SB_profiles}
\end{figure*}

SB profiles for the $g$, $r$, and $i$ bands, with the best-fitting convolved and deconvolved Sérsic models overlaid, are shown in \Fig\ref{fig:SB_profiles}.

\section{Database}

Some basic properties of the data sample—such as colours, absolute magnitudes, stellar masses, and structural parameters—are reported in Tables \ref{tab:tab1} and \ref{tab:tab2}.

\begin{table*}
\centering
\begin{threeparttable}
\caption{Parameters of dwarf candidates.}\label{tab:tab1}
%\resizebox{\textwidth}{!}{
\begin{tabular}{l c c c c c c c c c}
\hline
Name & RA & DEC & Vote & $e$ & PA & $g-r(R_{\rm e})$ & $r-i(R_{\rm e})$ & $M_r$ & $\log_{10}(M_\star/M_\odot)$ \\
 & hh:mm:ss & dd:mm:ss  &  &  & deg & mag & mag & mag & dex \\
\hline \\
dw1316-2107 & 13:16:46.6 & -21:07:04.9 & A & 0.18 & 124.9 & 0.68 & 0.23 & -14.94 & 8.42 \\
dw1316-2217 & 13:16:25.4 & -22:17:14.9 & B & 0.36 & 143.4 & 1.08 & 0.40 & -14.48 & 8.90 \\
dw1316-2218 & 13:16:55.8 & -22:18:20.4 & A & 0.20 & 71.0 & 0.45 & 0.36 & -11.47 & 6.65 \\
dw1317-2035 & 13:17:18.5 & -20:35:30.8 & A & 0.00 & 0.0 & 0.56 & 0.04 & -11.16 & 6.71 \\
dw1317-2118 & 13:17:46.5 & -21:18:40.3 & B & 0.11 & 132.6 & 0.54 & 0.24 & -12.21 & 7.09 \\
dw1317-2137 & 13:17:39.5 & -21:37:13.0 & B & 0.37 & 160.3 & 0.60 & 0.25 & -17.05 & 9.13 \\
dw1317-2142 & 13:17:43.9 & -21:42:33.5 & B & 0.72 & 4.2 & 0.29 & 0.10 & -12.16 & 6.67 \\
dw1317-2151 & 13:17:06.8 & -21:51:39.8 & B & 0.11 & 39.4 & 0.43 & 0.24 & -13.34 & 7.37 \\
dw1317-2209 & 13:17:19.8 & -22:09:24.0 & B & 0.00 & 0.0 & 0.47 & 0.67 & -10.11 & 6.14 \\
dw1317-2213 & 13:17:07.6 & -22:13:33.6 & B & 0.11 & 117.7 & 0.67 & 0.72 & -10.61 & 6.67 \\
dw1317-2223 & 13:17:48.4 & -22:23:37.1 & B & 0.11 & 10.8 & 0.57 & 0.13 & -11.29 & 6.78 \\
dw1317-2227 & 13:17:13.9 & -22:27:14.0 & A & 0.32 & 74.7 & 0.38 & 0.27 & -12.75 & 7.04 \\
dw1318-2041 & 13:18:30.5 & -20:41:17.0 & B & 0.26 & 153.7 & 0.37 & 0.17 & -16.75 & 8.64 \\
dw1318-2122 & 13:18:56.8 & -21:22:30.2 & B & 0.55 & 156.7 & 0.33 & 0.07 & -12.04 & 6.68 \\
dw1318-2134 & 13:18:17.4 & -21:34:59.2 & A & 0.57 & 40.7 & 0.59 & 0.22 & -12.29 & 7.21 \\
dw1318-2153 & 13:18:04.0 & -21:53:04.2 & A & 0.24 & 50.0 & 0.61 & 0.33 & -14.59 & 8.16 \\
dw1318-2201 & 13:18:36.3 & -22:01:40.0 & B & 0.34 & 105.0 & 0.68 & 0.34 & -12.18 & 7.32 \\
dw1318-2206 & 13:18:49.1 & -22:06:58.5 & A & 0.00 & 0.0 & nan & nan & -12.77 & nan \\
dw1319-2030 & 13:19:28.6 & -20:30:14.3 & A & 0.00 & 0.0 & 0.39 & 0.27 & -12.58 & 7.00 \\
dw1319-2041 & 13:19:28.9 & -20:41:35.3 & A & 0.12 & 164.3 & 0.62 & 0.29 & -12.47 & 7.33 \\
dw1319-2045 & 13:19:12.0 & -20:45:37.7 & A & 0.23 & 60.0 & nan & -0.08 & -12.60 & nan \\
dw1319-2117 & 13:19:16.6 & -21:17:07.0 & B & 0.51 & 71.6 & 0.86 & 0.30 & -12.85 & 7.88 \\
dw1319-2123 & 13:19:34.5 & -21:23:34.6 & B & 0.36 & 82.3 & 0.10 & 0.06 & -11.34 & 6.02 \\
dw1319-2134 & 13:19:44.1 & -21:34:13.7 & B & 0.12 & 174.7 & 0.60 & 0.28 & -11.50 & 6.91 \\
dw1319-2143 & 13:19:16.9 & -21:43:24.5 & A & 0.00 & 0.0 & 0.64 & 0.43 & -10.97 & 6.77 \\
dw1319-2149A & 13:19:02.1 & -21:49:34.2 & A & 0.47 & 90.0 & 0.58 & 0.21 & -12.42 & 7.24 \\
dw1319-2149B & 13:19:35.1 & -21:49:55.2 & B & 0.17 & 68.4 & 0.45 & 0.17 & -11.40 & 6.62 \\
dw1320-2027 & 13:20:30.6 & -20:27:27.4 & A & 0.23 & 24.3 & 0.45 & 0.00 & -14.90 & 8.03 \\
dw1320-2031 & 13:20:56.9 & -20:31:17.3 & A & 0.05 & 153.8 & 0.63 & 0.30 & -13.19 & 7.63 \\
dw1320-2032A & 13:20:12.2 & -20:32:48.4 & A & 0.43 & 131.1 & nan & nan & -11.99 & nan \\
dw1320-2032B & 13:20:27.6 & -20:32:26.9 & B & 0.00 & 0.0 & 0.57 & 0.05 & -12.11 & 7.10 \\
dw1320-2035 & 13:20:46.8 & -20:35:45.7 & A & 0.60 & 15.0 & 0.58 & 0.39 & -12.86 & 7.43 \\
dw1320-2036 & 13:20:08.5 & -20:36:11.5 & A & 0.39 & 84.1 & 0.58 & 0.19 & -13.13 & 7.53 \\
dw1320-2044 & 13:20:02.6 & -20:44:50.4 & A & 0.32 & 122.3 & 0.53 & 0.33 & -13.72 & 7.69 \\
dw1320-2108 & 13:20:45.4 & -21:08:50.1 & A & 0.00 & 0.0 & 0.62 & 0.12 & -11.11 & 6.79 \\
dw1320-2130 & 13:20:31.9 & -21:30:33.1 & A & 0.60 & 115.0 & 0.65 & 0.25 & -12.45 & 7.38 \\
dw1320-2135 & 13:20:42.9 & -21:35:44.5 & B & 0.22 & 22.6 & 0.48 & 0.29 & -11.51 & 6.72 \\
dw1320-2147 & 13:20:07.3 & -21:47:20.8 & A & 0.23 & 112.9 & 0.52 & 0.30 & -13.86 & 7.72 \\
dw1320-2200 & 13:20:12.6 & -22:00:09.3 & A & 0.26 & 176.2 & 0.22 & 0.06 & -14.39 & 7.44 \\
dw1320-2202 & 13:20:25.8 & -22:02:04.0 & A & 0.30 & 41.8 & 0.63 & 0.23 & -12.51 & 7.37 \\
dw1320-2211 & 13:20:16.0 & -22:11:14.9 & A & 0.00 & 0.0 & 0.46 & 0.19 & -12.24 & 6.97 \\
dw1320-2220 & 13:20:11.6 & -22:20:41.8 & A & 0.30 & 26.3 & 0.58 & 0.37 & -12.63 & 7.32 \\
dw1321-2100 & 13:21:27.5 & -21:00:27.6 & A & 0.17 & 121.7 & 0.25 & 0.37 & -10.88 & 6.09 \\
dw1321-2108 & 13:21:27.4 & -21:08:35.9 & A & 0.27 & 73.7 & 0.63 & 0.17 & -12.89 & 7.53 \\
dw1321-2113A & 13:21:15.3 & -21:13:15.6 & A & 0.28 & 90.0 & 0.72 & 0.14 & -12.42 & 7.48 \\
dw1321-2113B & 13:21:44.9 & -21:13:29.0 & A & 0.34 & 92.5 & 0.34 & 0.18 & -13.46 & 7.26 \\
dw1321-2201 & 13:21:04.0 & -22:01:57.9 & A & 0.17 & 94.9 & 0.60 & 0.33 & -15.42 & 8.48 \\
\hline
\end{tabular}
\begin{tablenotes}
\item Notes: Magnitudes and masses are calculated assuming a distance of 26 Mpc. From left to right, columns report: 
    a) galaxy name, 
    b) Right Ascension, 
    c) Declination, 
    d) classification vote, 
    e) ellipticity $e$,
    f) position angle (PA) in degrees, measured counterclockwise from west,
    g) $g-r$ colour within \Rer, 
    h) $r-i$ colour within \Rer,, 
    i) absolute $r$-band total magnitude, 
    l) total stellar mass.
    \end{tablenotes}
    \end{threeparttable}
\end{table*}

\begin{table*}
\centering
\resizebox{\textwidth}{!}{
\begin{threeparttable}
\caption{Sérsic parameters and their uncertainties in the three bands.}\label{tab:tab2}
\begin{tabular}{l c c c c c c c c c c c c c c c c c c}
\hline
Name & $\mu_e$ (g,r,i) & $\sigma_{\mu_e}$ (g,r,i) & $R_e$ (g,r,i) & $\sigma_{R_e}$ (g,r,i) & n (g,r,i) & $\sigma_n$ (g,r,i) & magtot (g,r,i) \\
 & mag arcsec$^{-2}$ & mag arcsec$^{-2}$ & arcsec & arcsec &  &  &   mag \\
\hline
dw1316-2107 & 25.13, 24.61, 24.16 & 0.05, 0.03, 0.05 & 8.17, 8.85, 8.01 & 0.22, 0.16, 0.21 & 0.97, 1.09, 0.89 & 0.03, 0.02, 0.03 & 17.89, 17.14, 17.00 \\
dw1316-2217 & 25.21, 24.20, 24.04 & 0.08, 0.03, 0.05 & 5.79, 6.14, 7.45 & 0.24, 0.10, 0.22 & 0.92, 0.94, 1.05 & 0.06, 0.02, 0.04 & 18.73, 17.59, 16.96 \\
dw1316-2218 & 25.72, 25.28, 25.09 & 0.16, 0.12, 0.14 & 2.59, 2.55, 2.88 & 0.21, 0.17, 0.22 & 0.83, 0.89, 0.95 & 0.11, 0.09, 0.10 & 21.04, 20.60, 20.12 \\
dw1317-2035 & 27.67, 26.29, 26.88 & 1.04, 0.16, 0.95 & 7.15, 3.81, 6.51 & 5.86, 0.30, 4.84 & 1.04, 0.59, 0.93 & 0.60, 0.10, 0.56 & 20.68, 20.92, 20.15 \\
dw1317-2118 & 25.66, 25.06, 24.93 & 0.15, 0.08, 0.14 & 3.25, 3.04, 3.34 & 0.28, 0.14, 0.26 & 1.12, 1.18, 1.18 & 0.11, 0.06, 0.10 & 20.36, 19.87, 19.54 \\
dw1317-2137 & 24.73, 24.03, 23.74 & 0.04, 0.03, 0.03 & 19.72, 18.24, 17.28 & 0.48, 0.26, 0.31 & 1.07, 1.02, 1.02 & 0.03, 0.02, 0.02 & 15.53, 15.02, 14.84 \\
dw1317-2142 & 25.24, 24.79, 24.76 & 0.06, 0.05, 0.08 & 3.30, 3.11, 3.37 & 0.09, 0.07, 0.13 & 0.63, 0.51, 0.53 & 0.04, 0.03, 0.05 & 20.15, 19.91, 19.69 \\
dw1317-2151 & 24.59, 24.21, 24.11 & 0.05, 0.04, 0.06 & 3.73, 3.67, 3.99 & 0.10, 0.08, 0.14 & 0.80, 0.89, 0.99 & 0.04, 0.03, 0.04 & 19.14, 18.74, 18.41 \\
dw1317-2209 & 26.30, 26.17, 26.18 & 0.24, 0.27, 0.64 & 2.14, 2.17, 3.26 & 0.30, 0.31, 1.37 & 0.27, 0.67, 1.19 & 0.14, 0.18, 0.41 & 22.39, 21.96, 20.84 \\
dw1317-2213 & 25.98, 25.26, 24.50 & 0.24, 0.14, 0.13 & 2.03, 1.85, 1.80 & 0.23, 0.12, 0.10 & 0.51, 0.57, 0.53 & 0.17, 0.10, 0.09 & 22.03, 21.47, 20.80 \\
dw1317-2223 & 25.21, 24.69, 26.35 & 0.12, 0.09, 0.44 & 1.88, 1.85, 5.13 & 0.09, 0.08, 1.36 & 0.61, 0.74, 2.24 & 0.08, 0.07, 0.33 & 21.36, 20.79, 19.71 \\
dw1317-2227 & 24.67, 24.37, 24.66 & 0.09, 0.08, 0.15 & 2.99, 3.16, 4.52 & 0.13, 0.13, 0.37 & 0.65, 0.71, 1.06 & 0.06, 0.05, 0.11 & 19.78, 19.33, 18.66 \\
dw1318-2041 & 24.44, 23.99, 23.87 & 0.03, 0.02, 0.03 & 17.01, 16.49, 17.19 & 0.27, 0.18, 0.26 & 0.86, 0.78, 0.79 & 0.03, 0.02, 0.03 & 15.65, 15.32, 15.10 \\
dw1318-2122 & 25.34, 24.96, 24.75 & 0.09, 0.07, 0.12 & 2.94, 2.75, 2.58 & 0.14, 0.10, 0.16 & 1.01, 1.06, 0.92 & 0.06, 0.05, 0.08 & 20.30, 20.04, 20.03 \\
dw1318-2134 & 25.95, 25.26, 25.13 & 0.11, 0.07, 0.11 & 4.33, 4.02, 4.21 & 0.25, 0.13, 0.23 & 0.63, 0.57, 0.63 & 0.09, 0.05, 0.09 & 20.27, 19.79, 19.51 \\
dw1318-2153 & 26.27, 25.59, 25.37 & 0.09, 0.06, 0.08 & 13.34, 12.70, 13.64 & 0.62, 0.39, 0.61 & 0.82, 0.78, 0.85 & 0.06, 0.04, 0.05 & 18.04, 17.49, 17.08 \\
dw1318-2201 & 24.97, 24.31, 24.29 & 0.12, 0.07, 0.15 & 2.26, 2.22, 2.59 & 0.14, 0.08, 0.22 & 0.85, 0.96, 1.22 & 0.08, 0.05, 0.10 & 20.58, 19.90, 19.43 \\
dw1318-2206 & 27.88, 28.51, 29.75 & 0.44, 1.29, 3.84 & 8.62, 17.57, 92.00 & 2.45, 18.66, 321.25 & 1.17, 1.82, 2.78 & 0.29, 0.76, 2.18 & 20.43, 19.30, 16.73 \\
dw1319-2030 & 26.25, 26.05, 26.33 & 0.13, 0.11, 0.20 & 6.13, 6.54, 9.09 & 0.40, 0.37, 1.04 & 0.42, 0.60, 0.97 & 0.09, 0.08, 0.15 & 19.96, 19.49, 18.85 \\
dw1319-2041 & 26.82, 26.25, 26.30 & 0.19, 0.10, 0.17 & 6.14, 6.31, 8.33 & 0.64, 0.34, 0.85 & 0.88, 0.89, 1.03 & 0.13, 0.07, 0.12 & 20.24, 19.61, 18.99 \\
dw1319-2045 & 27.47, 27.10, 26.46 & 0.49, 0.20, 0.26 & 6.25, 9.44, 6.31 & 2.07, 1.11, 1.02 & 0.99, 1.11, 0.80 & 0.31, 0.13, 0.18 & 20.79, 19.48, 19.86 \\
dw1319-2117 & 25.95, 25.71, 25.25 & 0.12, 0.08, 0.09 & 3.55, 5.26, 4.81 & 0.23, 0.22, 0.24 & 1.06, 1.45, 1.25 & 0.09, 0.06, 0.07 & 20.47, 19.23, 19.04 \\
dw1319-2123 & 25.33, 25.30, 25.37 & 0.09, 0.07, 0.12 & 2.53, 2.68, 2.98 & 0.09, 0.08, 0.17 & 0.49, 0.51, 0.57 & 0.05, 0.04, 0.08 & 20.91, 20.74, 20.54 \\
dw1319-2134 & 25.44, 24.91, 24.56 & 0.12, 0.08, 0.14 & 2.08, 2.08, 1.99 & 0.12, 0.08, 0.14 & 0.93, 1.12, 1.05 & 0.09, 0.06, 0.10 & 21.18, 20.58, 20.34 \\
dw1319-2143 & 26.56, 26.20, 25.47 & 0.30, 0.20, 0.17 & 2.75, 3.19, 2.84 & 0.45, 0.37, 0.23 & 0.57, 0.76, 0.49 & 0.20, 0.13, 0.11 & 21.91, 21.11, 20.81 \\
dw1319-2149A & 26.62, 26.14, 25.99 & 0.16, 0.10, 0.15 & 5.71, 6.12, 6.28 & 0.46, 0.31, 0.50 & 0.69, 0.72, 0.79 & 0.11, 0.07, 0.12 & 20.31, 19.65, 19.41 \\
dw1319-2149B & 24.95, 24.46, 24.52 & 0.10, 0.08, 0.16 & 1.91, 1.74, 1.92 & 0.08, 0.07, 0.16 & 0.70, 0.79, 0.98 & 0.07, 0.06, 0.11 & 21.02, 20.68, 20.41 \\
dw1320-2027 & 24.88, 24.50, 24.54 & 0.04, 0.03, 0.05 & 9.33, 9.80, 9.76 & 0.17, 0.13, 0.22 & 0.41, 0.45, 0.50 & 0.03, 0.02, 0.03 & 17.69, 17.17, 17.18 \\
dw1320-2031 & 26.00, 25.54, 25.29 & 0.14, 0.09, 0.13 & 6.00, 6.48, 7.06 & 0.43, 0.33, 0.50 & 0.66, 0.79, 0.78 & 0.10, 0.07, 0.09 & 19.60, 18.89, 18.46 \\
dw1320-2032A & 26.90, 26.83, 27.24 & 0.65, 0.26, 1.49 & 5.41, 6.30, 7.49 & 2.15, 0.88, 7.19 & 1.24, 1.13, 1.55 & 0.50, 0.22, 1.15 & 20.43, 20.08, 19.96 \\
dw1320-2032B & 27.07, 26.66, 26.92 & 0.36, 0.19, 0.43 & 5.91, 6.38, 7.90 & 1.31, 0.72, 2.17 & 0.82, 0.95, 1.19 & 0.23, 0.13, 0.29 & 20.61, 19.97, 19.65 \\
dw1320-2035 & 26.69, 26.07, 26.31 & 0.17, 0.08, 0.15 & 8.42, 7.96, 12.04 & 0.79, 0.28, 1.01 & 0.45, 0.42, 0.85 & 0.10, 0.05, 0.10 & 19.69, 19.21, 18.29 \\
dw1320-2036 & 26.29, 25.69, 26.04 & 0.12, 0.07, 0.27 & 8.00, 7.52, 9.70 & 0.50, 0.27, 1.70 & 0.44, 0.43, 0.86 & 0.08, 0.05, 0.18 & 19.42, 18.95, 18.47 \\
dw1320-2044 & 25.97, 25.22, 25.21 & 0.08, 0.04, 0.07 & 8.70, 7.37, 9.75 & 0.36, 0.15, 0.36 & 0.81, 0.68, 0.82 & 0.06, 0.03, 0.05 & 18.67, 18.35, 17.66 \\
dw1320-2108 & 26.30, 25.89, 27.03 & 0.24, 0.12, 0.77 & 2.71, 3.04, 6.85 & 0.33, 0.18, 3.75 & 0.52, 0.65, 1.57 & 0.16, 0.08, 0.49 & 21.72, 20.96, 19.94 \\
dw1320-2130 & 27.04, 26.43, 26.25 & 0.19, 0.14, 0.15 & 7.36, 7.38, 8.54 & 0.77, 0.56, 0.62 & 0.53, 0.58, 0.55 & 0.12, 0.08, 0.09 & 20.28, 19.63, 19.15 \\
dw1320-2135 & 25.27, 24.77, 24.56 & 0.10, 0.08, 0.11 & 2.16, 2.07, 2.11 & 0.10, 0.08, 0.11 & 0.79, 0.85, 0.93 & 0.07, 0.05, 0.08 & 21.01, 20.57, 20.27 \\
dw1320-2147 & 25.14, 24.65, 24.32 & 0.06, 0.04, 0.05 & 5.57, 5.82, 5.81 & 0.16, 0.11, 0.15 & 0.84, 0.82, 0.78 & 0.04, 0.03, 0.03 & 18.79, 18.22, 17.91 \\
dw1320-2200 & 24.69, 24.52, 24.56 & 0.04, 0.03, 0.06 & 7.00, 7.26, 7.60 & 0.14, 0.11, 0.23 & 0.66, 0.69, 0.77 & 0.03, 0.02, 0.04 & 17.95, 17.68, 17.58 \\
dw1320-2202 & 26.11, 25.70, 25.34 & 0.11, 0.07, 0.11 & 4.44, 5.17, 4.65 & 0.22, 0.18, 0.26 & 0.63, 0.76, 0.69 & 0.07, 0.05, 0.07 & 20.38, 19.56, 19.47 \\
dw1320-2211 & 27.33, 26.79, 26.37 & 0.41, 0.12, 0.26 & 8.49, 8.06, 7.15 & 2.29, 0.47, 1.08 & 0.57, 0.52, 0.39 & 0.24, 0.08, 0.16 & 20.23, 19.84, 19.77 \\
dw1320-2220 & 26.49, 25.53, 25.47 & 0.19, 0.08, 0.15 & 6.50, 5.25, 6.25 & 0.70, 0.20, 0.53 & 0.91, 0.61, 0.85 & 0.13, 0.05, 0.11 & 19.77, 19.45, 18.87 \\
dw1321-2100 & 25.74, 25.65, 25.12 & 0.31, 0.20, 0.31 & 2.23, 2.29, 2.34 & 0.37, 0.24, 0.39 & 0.71, 0.94, 0.69 & 0.21, 0.14, 0.20 & 21.45, 21.19, 20.75 \\
dw1321-2108 & 26.93, 26.35, 26.22 & 0.28, 0.13, 0.30 & 7.45, 7.79, 8.40 & 1.30, 0.61, 1.62 & 1.06, 1.02, 1.04 & 0.19, 0.09, 0.20 & 19.85, 19.18, 18.89 \\
dw1321-2113A & 26.41, 25.82, 25.72 & 0.14, 0.07, 0.13 & 5.04, 5.46, 5.63 & 0.34, 0.19, 0.36 & 0.52, 0.60, 0.63 & 0.09, 0.05, 0.08 & 20.48, 19.66, 19.48 \\
dw1321-2113B & 25.62, 24.72, 24.45 & 0.16, 0.08, 0.16 & 6.71, 4.58, 4.13 & 0.65, 0.20, 0.38 & 1.61, 1.25, 1.29 & 0.12, 0.06, 0.12 & 18.57, 18.61, 18.55 \\
dw1321-2201 & 25.29, 24.71, 24.42 & 0.05, 0.03, 0.04 & 12.84, 12.91, 13.73 & 0.27, 0.19, 0.27 & 0.62, 0.65, 0.65 & 0.03, 0.02, 0.03 & 17.26, 16.65, 16.23 \\
\hline
\end{tabular}
\begin{tablenotes}
    \item Notes: from left to right, columns report: 
    a) galaxy name, 
    b), c) effective SB ($\mu_e$) and 1$\sigma$ uncertainty
    ($\sigma_{\mu_e}$), 
    d), e) effective rradius (\Re) and 1$\sigma$ uncertainty
    ($\sigma_{R_{\rm e}}$),
    f), g) Sérsic index ($n$) and 1$\sigma$ uncertainty
    ($\sigma_{n}$),
    h) total Sérsic magnitude. The median relative errors for the effective radius are 7\%, 4\%, and 8\%, for the Sérsic index are 14\%, 8\%, and 11\%, and the median errors on the surface brightness are 0.12, 0.08, and 0.14~mag~arcsec$^{-2}$ in the $g$, $r$, and $i$ bands, respectively.
    \end{tablenotes}
    \end{threeparttable}}
\end{table*}

\section{Comparing parameter distributions}\label{app:comparisons_statistics}

Our sample of dwarf candidates is broadly consistent with that presented by \cite{Venhola2018FDSIV}. Starting from the colour--magnitude diagrams, the median values of $M_{\rm r}$ (although this depends on the distance assumptions adopted in this work) and the colours $g-r$, $r-i$, and $g-i$ are consistent within the scatter, with only small differences in the median values. In particular, for $M_{\rm r}$ the median and the $1\sigma$ percentiles are $-12.5^{+1.1}_{-2.0}$ mag for our sample versus $-12.6^{+1.4}_{-2.6}$  mag for the V+18 sample. Similarly, we find $0.57^{+0.07}_{-0.20}$ mag versus $0.56^{+0.09}_{-0.14}$ mag for $g-r$, $0.25^{+0.11}_{-0.13}$ mag versus $0.28^{+0.10}_{-0.10}$ mag for $r-i$, and $0.80^{+0.14}_{-0.21}$ mag versus $0.85^{+0.13}_{-0.22}$ mag for $g-i$.

To provide a more robust quantification of the similarity between the two distributions, we use the Kolmogorov--Smirnov (KS) test, the Earth Mover's Distance (EMD), and the Mann--Whitney U test. The KS test measures the maximum vertical distance between the cumulative distribution functions (CDFs) of two samples to determine whether they are drawn from the same underlying probability distribution. The EMD, derived from the Wasserstein distance, quantifies the physical `work' required to transform the parameter distribution of one sample into that of the other, thus providing a non-parametric measure of global dissimilarity. We compute this distance in one dimension. To assess the impact of stochastic sampling noise, we establish a significance threshold by performing 1\,000 intra-sample bootstrap iterations. An EMD is considered physically significant only if it exceeds the 95th percentile of the EMD values obtained by comparing the VST-SMASH sample against itself, ensuring that any reported discrepancy is not driven by small-number statistics. Finally, we also employ the Mann--Whitney U test, a non-parametric rank-sum test used to determine whether the median values of the dwarf galaxy properties differ significantly between the two samples.

Using the KS test, $M_{\rm r}$, $g-r$, and $g-i$ show good agreement, with $p > 0.05$, indicating no statistically significant difference, whereas $r-i$ yields $p < 0.05$, suggesting a mild -- although not visually evident in the plots -- but statistically significant offset between the two samples. Similar conclusions are obtained from the Mann--Whitney U test. Although the EMD remains within $0.5$ times of the standard deviation of the Venhola sample for all parameters, this ratio is larger for $r-i$. By performing the intra-sample bootstrap analysis, we also find that the EMD remains below the significance threshold, indicating that the possible minor discrepancies (e.g. in $r-i$) may arise from the large scatter and limited statistics of the VST-SMASH sample, or simply because the discrepancy is too small to be physically significant.

While colours are independent of the assumed distance, absolute magnitude is not. Therefore, we explore the impact of the distance assumption by performing the test for the values of 5.2 and 40 Mpc adopted in the paper. Not surprisingly, at these two distances (in particular at the distance of NGC 5068), the two distributions show significant differences.

Regarding the structural parameters, we find good consistency for $\log R_{\rm e}/ \rm kpc$ ($0.68^{+0.34}_{-0.38}$ vs $0.64^{+0.62}_{-0.32}$) and for the Sérsic index $n$ ($0.77^{+0.29}_{-0.19}$ vs $0.83^{+0.39}_{-0.20}$) between the VST-SMASH and V+18 samples. The statistical tests also indicate that the two samples are indistinguishable in terms of $R_{\rm e}$ and $n$. However, similarly to the absolute magnitude, these results for $\Re$ change when adopting different distance assumptions, with indications of significant differences emerging in that case.

\end{appendix}

\end{document}